\documentclass[fleqn,usenatbib]{mnras}
\usepackage{newtxtext,newtxmath}
\usepackage[T1]{fontenc}
\DeclareRobustCommand{\VAN}[3]{#2}
\let\VANthebibliography\thebibliography
\def\thebibliography{\DeclareRobustCommand{\VAN}[3]{##3}\VANthebibliography}

\usepackage{graphicx}	
\usepackage{amsmath}	
\usepackage{bm}         
\usepackage{float}      
\restylefloat{table}    
\usepackage{siunitx}    
\usepackage[capitalise]{cleveref}  
\usepackage{nccmath}   
\usepackage{notes2bib}

\title[Neural Networks for Planetary Collisions]{Residual Neural Networks for the Prediction of Planetary Collision Outcomes}
\author[P. M. Winter et al.]{
Philip M. Winter$^{1}$\thanks{E-mail: winter@ml.jku.at},
Christoph Burger$^{1}$\thanks{equal contribution},
Sebastian Lehner$^{2}$\footnotemark[2],
Johannes Kofler$^{2}$, \and
Thomas I. Maindl$^{3,4}$,
Christoph M. Sch\"afer$^{1}$
\\
$^{1}$Institut f\"ur Astronomie und Astrophysik, Eberhard Karls Universit\"at T\"ubingen, Auf der Morgenstelle 10, 72076 T\"ubingen, Germany\\
$^{2}$Johannes Kepler University Linz, Altenberger Straße 69, 4040 Linz, Austria\\
$^{3}$SDB Science-driven Business Ltd, 85 Faneromenis Avenue, Ria Court 46, Suite 301, 6025 Larnaca, Cyprus\\
$^{4}$Department of Astrophysics, University of Vienna, T\"urkenschanzstraße 17, 1180 Vienna, Austria
}

\date{Accepted 2022 September 29. Received 2022 September 23; in original form 2022 April 21 \\ \\ This is a pre-copyedited, author-produced PDF of an article accepted for publication in Monthly Notices of the Royal Astronomical Society following peer review.} 

\pubyear{2022}

\begin{document}
\label{firstpage}
\pagerange{\pageref{firstpage}--\pageref{lastpage}}
\maketitle

\begin{abstract}
Fast and accurate treatment of collisions in the context of modern $N$-body planet formation simulations remains a challenging task due to inherently complex collision processes. We aim to tackle this problem with machine learning (ML), in particular via residual neural networks. Our model is motivated by the underlying physical processes of the data-generating process and allows for flexible prediction of post-collision states. We demonstrate that our model outperforms commonly used collision handling methods such as perfect inelastic merging and feed-forward neural networks in both prediction accuracy and out-of-distribution generalization. Our model outperforms the current state of the art in 20/24 experiments. We provide a dataset that consists of 10164 Smooth Particle Hydrodynamics (SPH) simulations of pairwise planetary collisions. The dataset is specifically suited for ML research to improve computational aspects for collision treatment and for studying planetary collisions in general. We formulate the ML task as a multi-task regression problem, allowing simple, yet efficient training of ML models for collision treatment in an end-to-end manner. Our models can be easily integrated into existing $N$-body frameworks and can be used within our chosen parameter space of initial conditions, i.e. where similar-sized collisions during late-stage terrestrial planet formation typically occur.
\end{abstract}


\section{Introduction}
\subsection{Planet formation background}
Planet formation is inherently connected to collisions on all scales, from $\mu $m-sized dust grains up to planet-sized bodies. The precise mechanisms of early planetary growth generally depend on the current conditions in the protoplanetary disc and the amount and (dominant) size of available building blocks \citep[e.g.,][]{kokubo02, mcneil05, johansen17}. Particularly for terrestrial planets, our current understanding suggests that their final phase of accretion comprises growth via pairwise collisions of up to planet-sized bodies, lasting on the order of tens to hundreds of Myr \citep[e.g.,][]{kokubo98, chambers98, agnor99, chambers01, kokubo06}. This is supported by the long accretion times of terrestrial planets in the Solar System, as well as features like Mercury's high bulk density, Earth's large moon, or Mars' hemispheric dichotomy, all believed to be the consequences of large-scale collisions of roughly similar-sized bodies. Indirect evidence for such encounters has also been found in extrasolar systems \citep[e.g.,][]{wyatt16} in the form of observed infrared excess caused by warm dust, interpreted as collision debris. These large collision events are of particular interest as they shape the final characteristics of terrestrial planets, and likely contribute to the broad compositional diversity of observed low-mass exoplanets \citep{marcus09, marcus10, inamdar16, bonomo19}. This phase of planet formation naturally also leads to radial mixing of material and allows for (dynamical and collisional) transport of volatiles such as water to the inner parts of the system, and especially to potential planets forming in the habitable zone \citep{morbidelli00, izidoro13, obrien14, obrien18, haghighipour16, burger20}.

Modeling of this final phase of planet formation is typically based on $N$-body simulations, where mainly the gravitational interaction of hundreds to thousands of bodies is followed for up to few hundred Myr \citep[e.g.,][as some of the more recent work]{chambers13, fischer14, obrien14, quintana14, quintana16}. As planet formation models become more sophisticated and aim to study more than the most basic outcome quantities, collision modeling has to keep up in order to avoid systematic errors caused by too crude approximations of the underlying physics.

\subsection{The collision treatment problem}
\begin{figure}
 \includegraphics[width=\columnwidth]{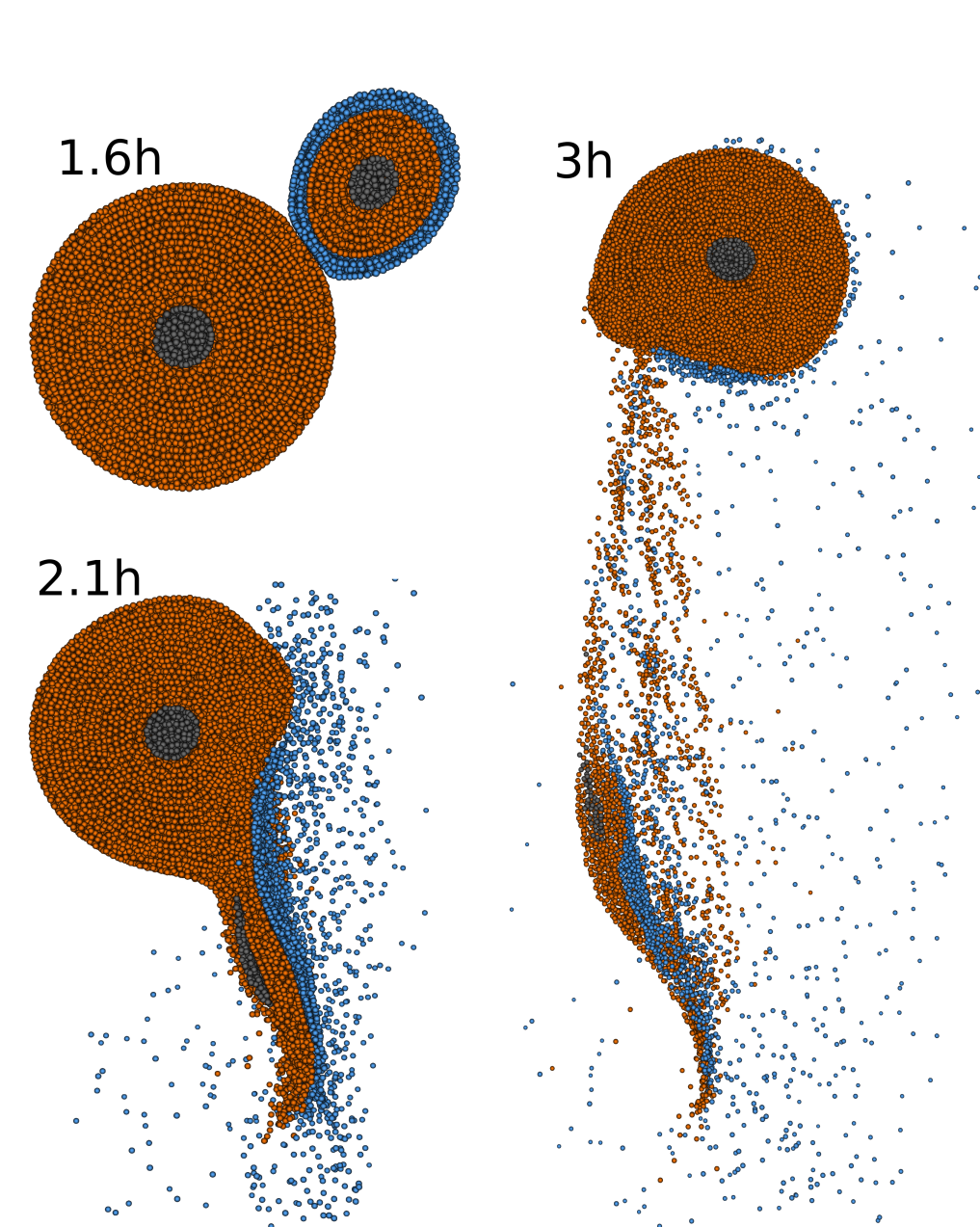}
 \caption{Exemplary snapshots at three different times of a SPH simulation of a planet-scale collision. The Mars-sized projectile hits the Earth-sized target at an impact angle of $43^{\circ}$ and an impact velocity of 1.3 times the mutual escape velocity, resulting in a hit-and-run outcome. Colors indicate the different materials -- an iron core, a silicate mantle, and a water/ice shell. Bodies are cut into halves for visualization. We perform 10164 collision simulations, covering a large parameter space of initial conditions.}
 \label{fig:sph_custom}
\end{figure}

Accurate modeling of major collisions among large, up to planet-sized bodies plays an important role in understanding the formation, evolution and diversity of planetary systems. The prediction task for two-body collisions is well-defined: Given the initial conditions such as collision geometry and object properties, we ask for the outcome state at a specific later point in time.

Up to relatively recently, collisions in planet formation scenarios were typically modeled by assuming complete accretion in all encounters \citep[e.g.,][]{haghighipour07, raymond04, raymond07, izidoro13, fischer14, obrien14, quintana14}, often referred to as perfect inelastic merging (PIM). This approach is simple and fast, but gives reasonably accurate predictions only for the lower end of the spectrum of characteristic collision velocities, or for large target-to-impactor mass ratios. In general, collisions between large and roughly similar-sized bodies can result in a diverse range of outcomes \citep[e.g.,][]{leinhardt12}, and often include significant material losses \citep{hagmai22}. This can affect bulk and chemical composition \citep[e.g.,][]{dwyer15, carter15, carter18}, and even more so for volatile constituents, especially at or close to the surface \citep{marcus10, maindl14, maindl17, burger18, burger20, kegerreis20}. In addition, collisions among similar-sized bodies frequently result in two large and gravitationally unbound survivors, instead of a single dominant one, as exemplified in \cref{fig:sph_custom}. These so-called hit-and-run events constitute up to half of all collision outcomes \citep[e.g.,][]{chambers13, clement19, burger20}. This can prolong planetary accretion considerably, naturally leads to a higher overall number of collisions, and can result in very different behavior in terms of material loss and transfer between colliding objects \citep{burger18, burger20proceedings, burger20}.

Several approaches have been developed to account for this diverse range of possible collision outcomes.
\citet{genda17a} developed scaling laws for collisional erosion with a focus towards smaller projectile-to-target mass ratios down to 1:10,000, where outcomes are generally dominated by a single large survivor.
\citet{zhou21} propose an approach that also exclusively assumes a single survivor, but includes randomly-picked material losses, based on statistics of a large number of SPH collision simulations.
\citet{crespi21} suggest an approach based on a catalogue of SPH collision outcomes, focusing on the distribution of smaller-scale collision fragments.
A recent framework based on semi-analytical scaling laws \citep{leinhardt12, stewart12, leinhardt15} has been applied in various planet formation studies \citep[e.g.,][]{chambers13, bonsor15, carter15, carter18, quintana16, clement19}. Albeit fast and relatively straight-forward to implement, its prediction accuracy for more complex behavior, like the fate of surface volatiles, or individual material losses and transfer in hit-and-run, is naturally limited \citep{burger18}.
\citet{genda11, genda17b} and \citet{burger20} resolve collisions in $N$-body planet formation simulations by running dedicated SPH simulations for each event on the fly, which is the most accurate approach, but computationally complex and expensive.

To summarize, depending on the problem at hand and the available computational resources, one has to make design choices which method to use. Both, simple problems and/or sufficient computational resources allow the use of sophisticated collision treatment methods, whereas complex problems and/or limited resources require certain trade-offs between prediction accuracy and computation time. For many applications it would be desirable to choose and adjust this trade-off more flexibly. Although analytic and heuristic approaches are efficient, they are typically neither very accurate, nor allow adjusting the accuracy-speed trade-off. In contrast, full hydrodynamic simulations for individual collisions are much more costly, but yet very accurate. In this paper we aim to combine all three properties, yielding an efficient, still accurate and flexible approach, where flexible means that it can be easily adapted to different accuracy-speed trade-offs.

\subsection{ML for planetary collisions}
The recent progress of cheap and efficient hardware caused a renaissance of ML, enabling to solve complex tasks in different fields such as computer vision \citep{krizhevsky17} and natural language processing \citep{brown20} with unprecedented accuracy and speed. Recently, \citet{tamayo20} and \citet{cranmer21} applied ML for predicting long-term stability and dissolution of compact multi-planet systems, indicating that ML may serve as an efficient tool for fast and accurate approximation of astrodynamical processes.

Recurrent neural networks \citep[RNNs;][]{jordan86, elman90, pearlmutter89} have been applied for approximating hydrodynamical simulations \citep{wiewel19} and astrophysical simulations such as 2D mantle convection \citep{agarwal21}. Several works successfully demonstrated the applicability and usefulness of ML for planetary collision treatment, opening up a promising research direction for computational astrophysics: \citet{valencia19} apply gradient boosting regression trees \citep{friedman01, breiman84}, Gaussian processes \citep[GPs;][]{rasmussen05}, and a nested method for classifying collision scenarios and regressing the largest remnant mass. \citet{cambioni19} use a multi-class support-vector machine \citep{cortes95, hearst98} for classification of different collision scenarios. They apply a small, 3-layered feed-forward neural network \citep[FFN;][]{rosenblatt61, ivakhnenko65} to regress accretion efficiencies, i.e. the mass of the largest remnant. \citet{cambioni21} extend this work and include surrogate models for predicting core mass-fractions of the largest and second-largest remnants. \citet{emsenhuber20} extend the work from \citet{cambioni19} to additionally predict orbital parameters of the two largest remnants with a separate regressor, resulting in a set of models that can be directly incorporated into $N$-body frameworks for collision treatment. However, this approach is limited to the main collision plane and does not allow prediction of orbital inclinations and longitudes of ascending nodes. The above mentioned works use the SPH data from \citet{reufer12} that consists of collisions between non-rotating, differentiated iron-silicate bodies.

\citet{timpe20} establish a high-quality dataset that consists of 14856 collisions between differentiated, rotating bodies \citep{timpe20_data}. They apply a two-step classification-regression approach to predict post-collision properties. They study several different methods for collision treatment and find data-driven methods to outperform non-data driven methods. Gradient boosted decision trees and FFNs are used for both classification and regression, whereas polynomial chaos expansion \citep{wiener38} and GPs are studied for regression only. They train different regressors for each individual post-impact property, and predict a variety of properties of the largest and second largest remnant, and the remaining debris. FFNs and XGBoost \citep{chen16} perform best amongst data-driven methods. We regard that study as our closest related work.

The overall goal of our work is to improve the prediction of planetary collision outcomes via ML models. In particular, this includes minimizing systematic prediction errors as much as possible by outperforming the current state of the art. We improve upon the works above by providing a more general dataset, reframing the ML task as a multi-task problem, and employing a simple, but problem-adapted ML model for the prediction of planetary collision outcomes. We train our model to predict masses, material fractions, positions, and velocities of the two largest post-collision remnants, and the remaining debris. Our contributions are summarized as follows:
\begin{itemize}
    \item We perform extensive $N$-body simulations to determine realistic initial conditions for planetary collisions. We base the choice of the parameter space for our SPH dataset on the outcome of the $N$-body simulations. To that end, we provide a comprehensive dataset that consists of 10164 SPH simulations of pairwise planetary collisions. We use between 20k to 50k SPH particles, which is relatively low-resolution compared to state-of-the-art simulations in astrophysics with up to several million SPH particles. Our dataset covers typical collision setups and is the first of its kind to combine all essential elements for a comprehensive treatment of collisions, including realistic object models (differentiated and rotating bodies), detailed pre- and post-collision geometries, and temporal information. The dataset allows to study several generic topics, such as collision treatment in a broad range of scenarios, inverse problems (e.g., the Moon-forming impact), and collisional accretion during planet formation. While our dataset is in general comparable to the one provided by \citet{timpe20}, it additionally includes volatile (water) layers, which opens up studies regarding collisional water/volatile transfer and loss, even though this is intended rather as a proof of concept in this work, mainly because of the difficulty to accurately resolve such surface layers.
    \item In contrast to existing work we follow a multi-task learning approach in the sense of multi-dimensional regression, in which a single ML model learns to predict the entire post-collision state rather than only specific, individual aspects of the state. Our ML task generalizes the collision treatment problem to 3D space, while at the same time avoiding the need for manual definition of class boundaries for different collision scenarios. Existing approaches often formulate the task as a classification problem, requiring somewhat arbitrary class definitions. We demonstrate that our multi-task learning approach leads to simple and computationally efficient models, while remaining relatively accurate compared to single-task learning.
    \item We propose an ML model which helps modeling of temporal dynamics by evolving system states in an autoregressive manner. This closely resembles the data generation process, i.e.\ classical numerical solvers that iteratively solve the underlying hydrodynamic equations. This includes handling both, the properties of the colliding bodies, as well as the spatio-temporal evolution of the system. Our model allows for flexible prediction of post-collision states at different times, and can be employed for collision treatment within existing $N$-body frameworks. We demonstrate superior prediction accuracy in comparison to commonly used baseline methods and the current state of the art. Moreover, our model requires little computational costs, reducing the prediction speed by approximately 4 orders of magnitude compared to the SPH simulations.
\end{itemize}

With our work we aim to provide high-quality data and an ML model that is useful for various downstream applications.
The paper is organized as follows. In Section 2, we describe our data generation pipeline, as well as the ML model used for collision treatment. In Section 3, we present our experiments and their results. Section 4 summarizes and concludes the paper.

\section{Methods}
\subsection{Data generation}
\subsubsection{$N$-body simulations}
\label{sect:nbody}
\citet{burger20} developed a hybrid framework, based on extensive $N$-body simulations in combination with realistic collision treatment by direct SPH simulations.
These results and collision statistics are also used to inform the choice of initial conditions for the SPH simulations performed in this study. In addition, we provide a cleaned and extended\footnote{Based on new (yet unpublished) $N$-body + SPH simulations in a similar dynamical environment.} version of their dataset of approximately 10k collisions, which we refer to as `$N$-body dataset'\footnote{The `$N$-body dataset' based on the simulations by \citet{burger20} provides data on collision parameters before contact, and basic data on the final state after the collision, like masses and composition of the two largest remnants, but no dynamical information (positions and velocities) and no data on intermediate states. Along with our other data and tools it is available at \url{https://github.com/littleblacksheep/csv/tree/main/misc}.}.

The scenarios in \citet{burger20} are based on an evolving disc of ($\sim$Mars-mass) planetary embryos + smaller bodies (planet\-esimals). Their dynamical and collisional evolution is followed over several hundred Myr of terrestrial planet formation in an environment akin to the early Solar System. The embryos and planetesimals are modeled as differentiated, three-layered, self-gravitating bodies, similar to the SPH simulations in this work. The rotation state is not tracked across multiple collisions. The approach of on-the-fly SPH simulations allows not only for accurate treatment of each individual collision, but also a relatively straight-forward re-integration of collision outcomes into the overall $N$-body dynamics (for our ML approaches this is discussed in \cref{sec:drop_in}). It also includes individual tracking of both large survivors in hit-and-run collisions, which comprise up to 50 per cent of outcomes between similar-sized bodies. Therefore this dataset also provides reliable collision (input parameter) statistics for the scenarios in this work.

\subsubsection{SPH simulations}
\label{sect:sph}
\begin{figure}
\begin{center}
 \includegraphics[width=0.8\columnwidth]{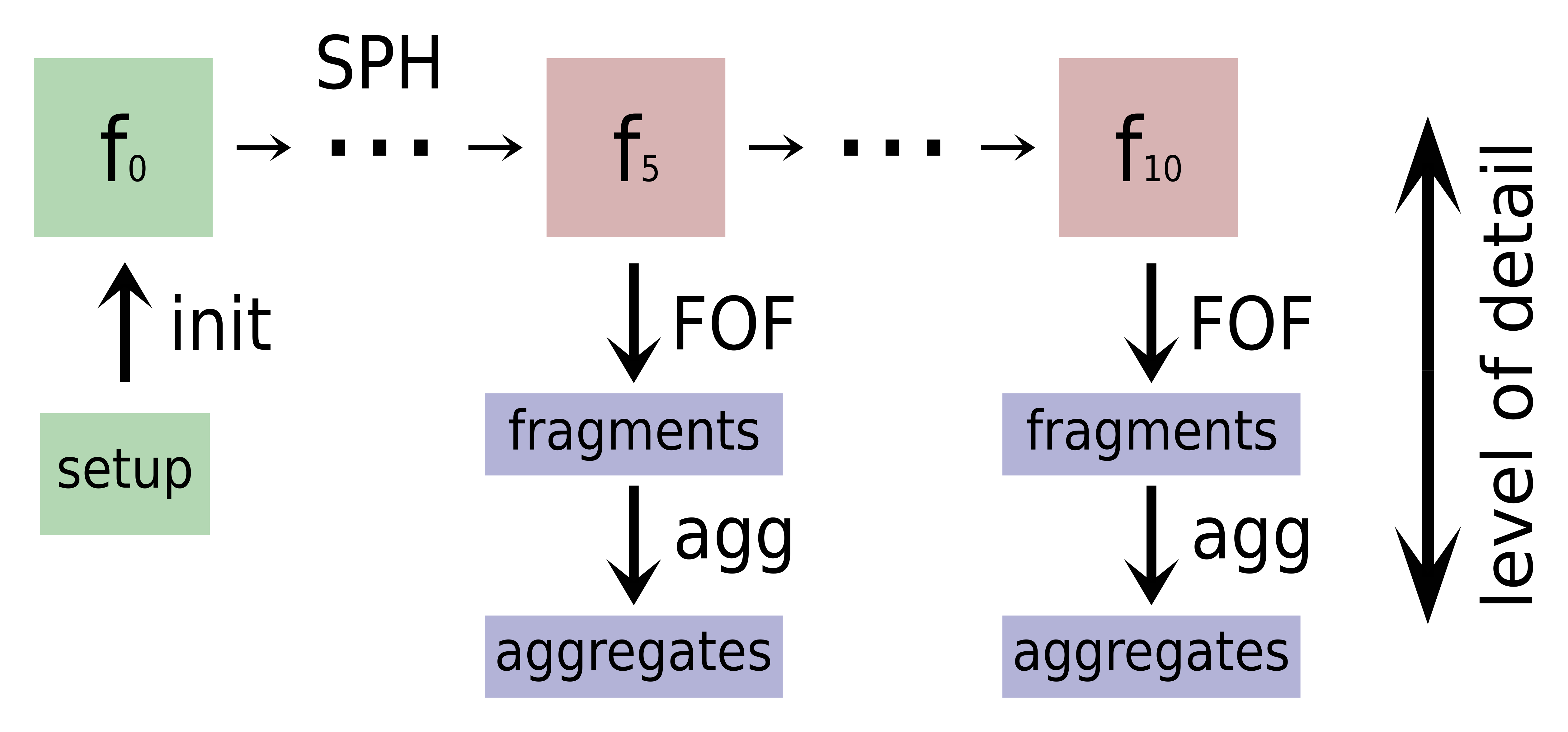}
\end{center}
 \caption{Data generation pipeline. For each individual simulation, a setup is sampled from the parameter space. The SPH particle distribution and its properties are set up in 'init', resulting in the input frame $f_0$. The SPH code then evolves the system, leading to a number of output frames. All output frames are postprocessed with a friends-of-friends algorithm (FOF) to compute all spatially connected material 'fragments'. Finally, 'aggregates' are identified, defined as gravitationally-bound collections of fragments.}
 \label{fig:data_pipeline}
\end{figure}

SPH is a numerical method for modeling visco-elastic fluid flows. The method was first proposed by \citet{monaghan77} and \citet{lucy77} and has since been applied extensively to model various aspects of astrophysical collision processes, including planetary collisions. In this work, we use the SPH code \texttt{miluphcuda}\footnote{The SPH code \texttt{miluphcuda} is in active development and publicly available at \url{github.com/christophmschaefer/miluphcuda}.} \citep{schaefer16, schaefer20} to generate a planetary collision dataset. An example is illustrated in \cref{fig:sph_custom}. The SPH code solves the continuum mechanics equations for hydrodynamic flow, can handle three dimensional, multi-material problems, and includes self-gravity. It also includes modules for the simulation of elasto-plastic solid-body physics based on several available material models and equations of state.

In this work, we perform pure hydro simulations, i.e., only solving the Euler equation with scalar pressure, instead of full tensorial treatment of material strength. Since we perform a large number of simulations, we trade some physical accuracy for numerical stability and more data (due to faster computation). However, this design choice is still a reasonably good proxy within the scope of our scenarios \citep{burger17, burger18}. For actual collisions in an active planet formation environment it can be assumed that the physical state of the colliding bodies -- and hence their material (strength) response -- varies over a broad range, even for otherwise identical scenarios in terms of masses, compositions, and collision parameters. This may be a function of their collision history, thermal state, and possibly various other factors. Considering those ambiguities, our rather simple material model allows the dataset to remain as general as possible and at the same time consistent over our whole parameter space. We use the Tillotson equation of state \citep{tillotson62, melosh89} for all simulations. Technical details are given in \cref{SPH_technical}.

\begin{figure}
\begin{center}
 \includegraphics[width=0.5\columnwidth]{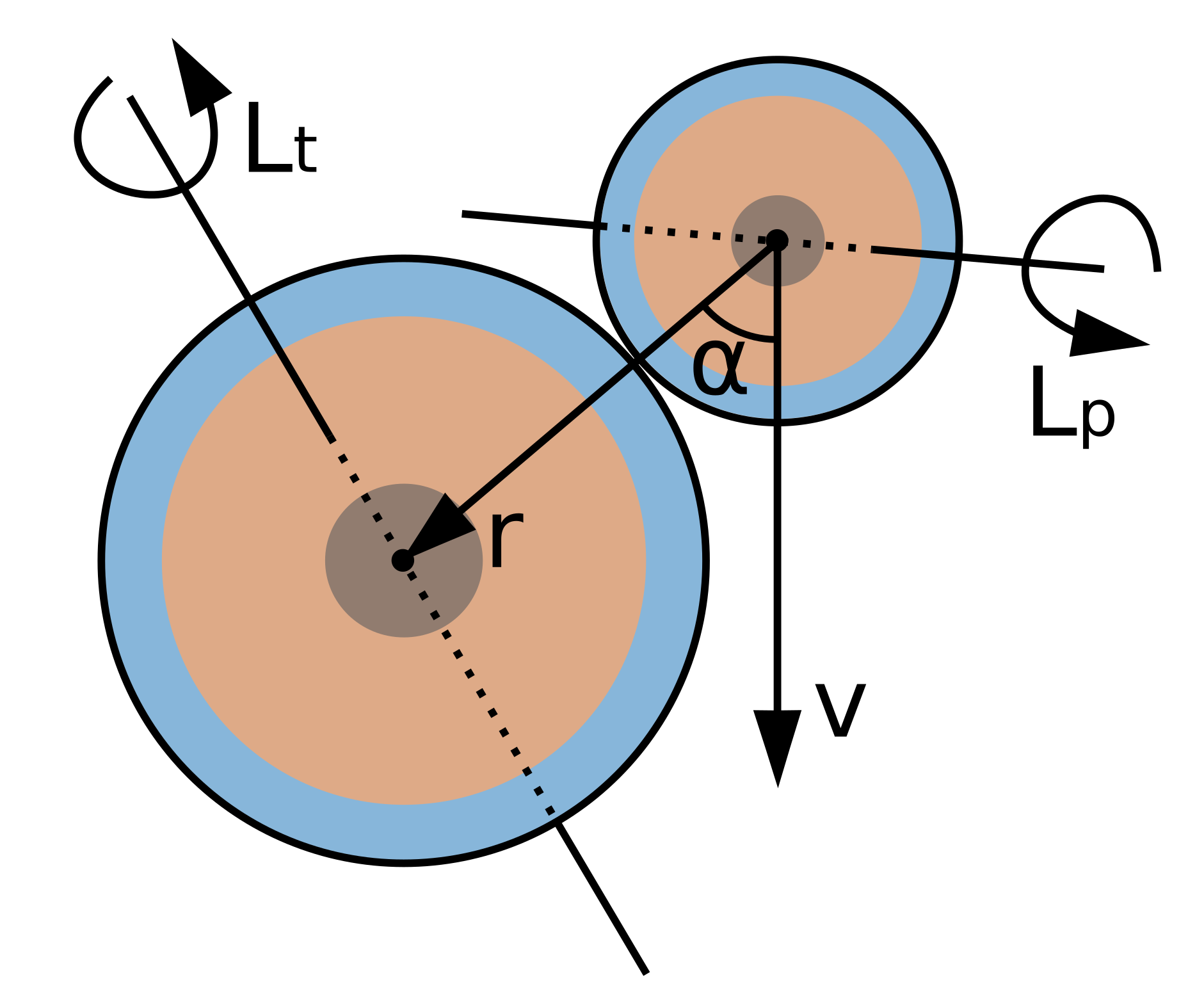}
 \end{center}
 \caption{Collision geometry for planetary collisions. The impact angle $\alpha$ is measured between the relative position \textbf{r} and the relative velocity \textbf{v} between target and projectile at 'touching-ball' distance. Both objects comprise a core-mantle-shell structure and have random rotation axes $\mathbf{L}_t$ and $\mathbf{L}_p$, which can lie outside the plane spanned by \textbf{r} and \textbf{v}.}
 \label{fig:collision_geometry}
\end{figure}

The SPH simulation pipeline is fully automated and includes all steps to initialize, run, and postprocess individual simulations (see \cref{fig:data_pipeline}). For each run, a specific parameter set is sampled from the parameter space (\cref{tab:param_space}). The chosen parameters cover a broad range of possible collision scenarios during terrestrial planet formation. The particular choices of parameter ranges are additionally informed by the robust statistics of our $N$-body dataset (see \cref{sect:nbody}). Note that we use the $N$-body dataset exclusively for choosing meaningful parameter intervals representative of late-stage terrestrial planet formation. For creating the SPH dataset, our parameter space of initial conditions is sampled randomly within the chosen intervals.

\begin{table}[ht]
\centering
\begin{tabular}{l|l|l|l} 
 parameter & min & max & description \\ 
 \hline
 $M_\mathrm{tot} [kg]$ & $2\times M_\mathrm{Ceres}$ & $2\times M_\mathrm{Earth}$ & total mass \\
 $\gamma [1]$ & 0.05 & 1 & mass ratio $m_p/m_t$ \\
 $\zeta_\mathrm{iron} [1]$ & 0.01 & 0.25 & iron (core) fraction \\
 $\zeta_\mathrm{water} [1]$ & 0; 0.1 & 0.25 & water (shell) fraction \\
 \hline
 $v_\mathrm{imp} [v_\mathrm{esc}]$ & 1 & 8 & impact velocity \\
 $\alpha [\mathrm{deg}]$ & 0 & 90 & impact angle \\
 \hline
 $P_\mathrm{rot} [P_\mathrm{rot, crit}]$ & 0 & 0.2 & rotation period \\
 $\theta_\mathrm{rot} [\mathrm{deg}]$ & 0 & 180 & rotation axis polar \\
 $\phi_\mathrm{rot} [\mathrm{deg}]$ & 0 & 360 & rotation axis azimuthal  \\
 \hline
 $f_i [1]$ & 3 & 7 & initial distance factor \\
 $f_t [1]$ & 40 & 60 & simulation time factor \\
 $N_\mathrm{tot} [1]$ & 20k & 50k & number of SPH particles
\end{tabular}
\caption{Parameter space of initial conditions for our SPH simulations, covering a wide range of typical scenarios for rocky planet formation. See the text for detailed definitions. All parameters are randomly sampled.}
\label{tab:param_space}
\end{table}

For initializing self-gravitating bodies in hydrostatic equilibrium we adopt the approaches and tools from \citet{burger18}, who calculate realistic density and pressure profiles for multi-layered bodies. The colliding objects are referred to as \textit{projectile} and \textit{target}, the latter being the more massive body. They are initialized at a certain distance, on the order of several times the sum of their radii, to allow for pre-collision tidal deformation, relaxation of rotating configurations, and settling of residual numerical artefacts (e.g., at material boundaries). Based on the desired impact velocity and impact angle at 'touching-ball' distance (cf.~\cref{fig:collision_geometry}), initial positions are calculated via backtracking the analytical two-body trajectories up to a distance of $d_\mathrm{initial} = f_i \times (R_t + R_p)$. $R_t$ and $R_p$ are the target and projectile radii, and the initial distance factor $f_i$ is a parameter. The total simulation time is calculated via $T_\mathrm{sim} = \tau_\mathrm{col} \times (f_{i} + f_{t})$ and rounded up to the next full hour. $\tau_\mathrm{col}$ is the collision time-scale $\tau_\mathrm{col}=(R_t + R_p)/v_\mathrm{imp}$. The impact velocity $v_\mathrm{imp}$ and the impact angle $\alpha$ are specified at touching-ball distance $R_t + R_p$, where $\alpha=0^{\circ}$ corresponds to head-on collisions and $v_\mathrm{imp}$ is the absolute value of the relative velocity vector $v$ at touching-ball distance (cf.~\cref{fig:collision_geometry}). The minimum number of SPH particles is set such that the resulting water shell has a thickness of at least 2 SPH particles at $\zeta_\mathrm{water}=0.1$ \citep{burger19_diss}. Note that this resolution may be too low to accurately simulate the water layers' response for a range of scenarios (parameter combinations). This can be particularly problematic for the 2nd-largest post-collision remnant, while reliable predictions are possible for the largest remnant already at resolutions similar to ours, as demonstrated by \citet{burger18}. Nevertheless, results for water mass fractions on post-collision remnants should be taken with a grain of salt, and consequently we consider our ML model predictions for this aspect rather a proof of concept and not generally accurate at this point. For other basic outcome properties on the other hand, like masses and kinematics of the two largest remnants, \citet{burger18} found resolution convergence for similar collision scenarios within 10\% for their 100k particles simulations. Our simulations contain either 2 or 3 materials, depending on $\zeta_\mathrm{water}$, where we remove the water shell if $\zeta_\mathrm{water}<0.1$ was sampled. The total colliding mass covers a range from $2\times M_\mathrm{Ceres}$ up to $2\times M_\mathrm{Earth}$. The mantle (basalt) mass-fraction is defined by $\zeta_\mathrm{basalt} = 1 - \zeta_\mathrm{iron} - \zeta_\mathrm{water}$. Since the hydrostatic initialization routine is based on non-rotating objects, we set our maximum rotation period $P_\mathrm{rot, max}=0.2\times P_\mathrm{rot, crit}$ for both target and projectile in order to avoid excessive initial oscillations and instabilities, which typically occur once $P_\mathrm{rot}$ approaches $P_\mathrm{rot, crit}$. The critical rotation period $P_\mathrm{rot, crit}$ is defined such that material at the surface of the (idealized spherical) body is weightless according to Kepler's 3rd law. Rotation axes are chosen randomly for both target and projectile. We refer to \cref{rotations} for more details.

During simulation, the SPH code periodically produces output frames, which contain the state of all SPH particles at the respective time. We keep the first, the last, and intermediate frames for postprocessing, where intermediate frames are saved at 5-hour intervals (simulated time). All frames undergo the same postprocessing procedure:
\begin{enumerate}
    \item Spatially connected collision fragments are calculated by a friends-of-friends algorithm \citep{geller83}.
    \item Barycentres, orbital angular momentum, and spin angular momentum are calculated for each fragment, as well as for the entire system.
    \item The two\footnote{Gravity-dominated collisions of roughly similar-sized bodies generally result in either none (if highly destructive), one, or two (in hit-and-run scenarios) large surviving bodies, along with orders-of-magnitude smaller debris.} largest aggregates of fragments are calculated. An aggregate is defined as a collection of gravitationally bound fragments, determined by an iterative procedure, which starts from the most-massive fragment as seed \citep[see][for details]{burger20}. In the remainder of the paper, these aggregates are referred to as 'remnants' for clarity.
    \item Basic visualization is done for the large fragments. A fragment is considered significant if it consists of at least 5 SPH particles.
    \item In this work we focus on the prediction of macroscopic system states, requiring information on the level of remnants only. Moreover, we aim to keep memory requirements of the final dataset low. Therefore, SPH output frames are sub-sampled, keeping one out of ten SPH particles.
\end{enumerate}

Keeping intermediate frames enables in-depth studies of temporal properties of the collision process. Moreover, they allow for the development of sophisticated ML models, i.e. models that not only predict the final state of the system, but the entire temporal evolution in detail. Note that since we sample our parameter space randomly, inputs to ML models do not require initial conditions that are similar to those in \citet{burger20}.

\subsection{Machine Learning for collision treatment}
From an ML perspective, the collision treatment task requires learning physical laws (e.g. conservation laws, material deformations, gravitational interactions, etc.) and handling the temporal evolution of the system (e.g. via time-series modeling). Various ML approaches can be applied in different contexts, mostly depending on which level of detail one is interested in. Therefore, we design our SPH dataset such that it can be used at different levels of detail. For example, one can use remnant or fragment information ('macro states') rather than SPH particle representations ('micro states') for learning certain aspects (e.g. predicting certain quantities such as the mass of the largest remnant or the thermal energy of the system). Depending on which level of detail ML is applied to, different aspects may be able to be learned more or less efficiently. In this work we focus on macro states because this setup is the most relevant one in order to incorporate ML models into $N$-body simulations for planet formation and evolution (see \cref{tab:mode_features} and \cref{sec:drop_in} for more details). In contrast, ML models operating on micro states may be a better choice if one is interested in studying details of the hydrodynamic flow and physical interactions in simulations such as SPH.

\subsubsection{Collision treatment as a multi-task regression problem}
\label{sec:regression}
Supervised learning is the task of selecting (learning) a specific model from a certain model class by using example input-target pairs. The difference between model outputs and desired target outputs results in an error, which is used to improve a model. We train our ML models in a supervised manner to predict several different quantities (mass, material fractions, position, and velocities) at once, which turns the problem into a multi-task problem. Our multi-task problem can be interpreted as a multi-dimensional regression problem of different physical quantities, since we use shared representations to predict different modalities. We motivate formulating and solving the problem as a multi-task problem due to inherent dependencies and correlations between the individual sub-tasks (e.g., trajectories of individual fragments highly depend on the overall mass distribution). Since all of our sub-tasks are highly correlated with each other, we hope that the multi-task setting supports generalization due to shared representations within ML models, acting as a form of regularization. Shared representations naturally allow for exploiting dependencies and correlations between different tasks, potentially improving the ML model's predictive performance. Note that in contrast to our multi-task setting, \citet[][]{cambioni19, cambioni21, emsenhuber20, timpe20} use individual ML models for predicting either single or sub-sets of collision outcome quantities, i.e. following single-task approaches. We believe that single-task approaches introduce unnecessary restrictions to the generalization capabilities of ML models, because individual models are exclusively able to specialize in their respective regime. Also, using individual regression-models for individual outcome scenarios (i.e. erosion, accretion, and hit-and-run) may lead to various issues caused by data-scarcity due to class-imbalances, which are common in planetary collision datasets. Moreover, using a single ML model for solving several sub-tasks at once may allow much better accuracy-speed trade-offs, especially in the presence of many sub-tasks. In our work, the importance of the individual sub-tasks are implicitly given via data preprocessing, effectively weighting loss terms of the respective sub-tasks. However, the importance can be explicitly adjusted depending on specific use cases.

To our knowledge, this is the first work to fully formulate the ML task as a regression problem. Our formulation allows simple, yet efficient training in an end-to-end manner and facilitates easy integration of ML models into existing $N$-body frameworks. Our regression objective does not explicitly optimize for classification performance, but rather for regression of macroscopic properties of the system. At the same time, our objective avoids the need for complicated approaches which require two separate ML models for regression and classification, respectively.

We believe that it is not beneficial to train classifiers which explicitly discriminate between different outcome scenarios such as accretion or hit-and-run, because such classification can be easily performed as a postprocessing step on top of regression-model predictions. We favor performing classification via a postprocessing step rather than training dedicated classifiers because the former can be used in combination with variable class definitions, whereas the latter is bound to fixed class definitions. When training dedicated classifiers, changing class definitions would require re-training the classifiers, which can be quite cumbersome in practice. Moreover, pure classifiers can not be used as full replacement for collision treatment in $N$-body simulations.

In general, we believe that defining and learning fixed classification schemes is not optimal due to continuous transitions between classes and the associated arbitrariness of class definitions. We believe that training dedicated classifiers is only reasonable if one is explicitly interested in accurate classification under the restriction of fixed class definitions.

The integration of ML models for collision treatment into $N$-body simulations might require additional postprocessing steps on top of ML predictions. This includes restricting predictions to conserve certain quantities such as the total mass, e.g.~by re-scaling predicted masses and/or distributing debris material across the two largest remnants. For the actual application in $N$-body simulations it is especially important how (if at all) the remaining collision debris is treated, which typically consists mostly of physically non-connected and gravitationally unbound fragments. This naturally opens up many possibilities depending on the respective use case (i.e.~the precise physical and numerical model). In our experiments, we do not apply any additional postprocessing steps in order to remain as general as possible and to obtain conservative performance estimates.

Fragments that are formed from collisions between non-rotating objects mostly remain in the collision's main symmetry plane (the x-y plane in our case) with only marginal z-components. However, collisions between rotating objects generally break this symmetry, and may produce large fragments with significant z-components. This symmetry breaking is also confirmed by the data from \citet{timpe20}. We thus generalize the prediction task from \citet{emsenhuber20} to three-dimensional space, treating all dimensions equally to consistently handle deviations from the main collision plane.

\subsubsection{Autoregressive ML models for temporal evolution}
\label{sec:autoregressive_models}
The use of autoregressive ML models for predicting collision outcomes can be motivated by studying the data generation process, i.e., the SPH simulations. We know that the data generation process has the Markov property, i.e., states $s_{t+h}$ at a time $t+h$ depend entirely on their previous states $s_t$ at time $t$. We assume continuous transitions between states for infinitesimal stepsizes h. The transition from $s_t$ to $s_{t+h}$ is described by a transition function $g$.
\begin{ceqn}
\begin{equation}
    s_{t+h} = g(s_t, h)
\end{equation}
\end{ceqn}
Historically, $g$ refers to a set of hand-crafted equations that incorporate certain physical laws (e.g., gravity, friction, etc.), as well as a procedure to evolve the system in time (e.g., numerical integration) by means of differential equations. In practice these approaches often suffer from limitations, such as the requirement to use small stepsizes when using classical solvers. Too large stepsizes typically introduce large systematic errors, often leading to diverging or unstable solutions.

In this work we aim to approximate solutions obtained using hand-crafted transition functions via an ML model that is learned from data. We believe that ML models are -- once trained -- efficient, powerful, and flexible transition functions for modeling the underlying physical processes in planetary collisions over time. In contrast to FFNs, our proposed model class exploits the Markov property of the data generation process, i.e., taking multiple, autoregressive steps to predict system states at a desired time $T$. There are several arguments that support the use of autoregressive ML models:
\begin{itemize}
    \item Neural networks are universal function approximators \citep{hornik89} which allow learning highly complex functions. This property allows direct prediction of system states at various times $T$, entirely circumventing the need for time-series modeling. ML models should thus be much more computationally efficient compared to numerical simulations. The most extreme case would be to predict the final state directly, as typically achieved in the literature. Depending on the choice for the stepsize, our model allows a flexible accuracy-speed trade-off. Small stepsizes can be expected to better model the physical processes and lead to more accurate predictions at the cost of computational resources, whereas large stepsizes lead to less accurate, but faster predictions.
    \item Autoregressive models subdivide the prediction of system states by taking multiple iterative steps. Since the universal function approximation theorem also applies to autoregressive ML models, they can use magnitudes larger, more complex time-steps compared to classical transition functions (numerical solvers) before getting unstable. This property typically makes these ML models much more efficient in terms of computational costs compared to hand-crafted transition functions.
    \item Learned transition functions allow context-dependent time-steps, i.e., adjusting the transition function automatically, based on data-specific information. This property avoids algorithmic design decisions, making ML-based transition functions more general and flexible compared to hand-crafted transition functions.
    \item Using autoregressive ML models allows for improved interpretability by enabling analysis of intermediate states. Such an analysis is not possible for ML models like FFNs or regression trees, which typically try to predict final post-collision states directly.
    \item Due to their design, we believe that autoregressive ML models can achieve better generalization compared to methods that try to predict final states directly, allowing more accurate predictions and improved out-of-distribution generalization capabilities. Autoregressive ML models that learn fixed time intervals (i.e., taking multiple steps with the same stepsize) are effectively time-invariant per design, in the sense that they have to learn physical processes at only a single timescale. Thereby, the models are not forced to spend their parameters\footnote{Note that herein term "parameters" can either refer to individual parameters of initial conditions for our SPH simulations, or to learnable parameters of an ML model. Both cases should be apparent from the respective text passages.} for learning to become time-invariant. This property can potentially also lead to improved parameter efficiency compared to non-time-invariant ML approaches.
    \item Longer physical interactions typically lead to the emergence of more complex dynamical processes during planetary collision events. Using autoregressive models naturally accounts for these effects by allocating computational resources that linearly scale with time, which is consistent with fluid-flow approaches.
\end{itemize}

Gated architectures \citep{hochreiter97, kyunghyun14} and regularized RNNs \citep{schmidt21} are able to produce chaotic dynamics, but often suffer from the exploding gradient problem \citep{metz21}, typically leading to diverging sequences \citep{monfared21}. On the other hand, non-chaotic sequences have bounded loss gradients and converge to fixed points. Thus, training autoregressive ML models is typically non-trivial and often very sensitive to hyperparameters, especially when the data-generating process is itself chaotic. Exploding gradients and diverging sequences in LSTMs\footnote{Long Short-Term Memory, a special kind of RNN that is widely used for processing sequential data such as written texts, time series, or DNA sequences.} can be mitigated via the forget gate \citep{gers99}, thereby reintroducing the vanishing gradient problem \citep{hochreiter98}. The vanishing gradient problem can prohibit efficient training of deep neural networks. For our autoregressive ML model, we find regularization of hidden states to be a robust strategy against diverging sequences. Moreover, we find that gradient descend with backpropagation through time \citep{mozer95, robinson87, werbos88} works fine for training.

\subsubsection{Residual neural network for planetary collision handling}
\begin{figure}
\begin{center}
 \includegraphics[width=0.8\columnwidth]{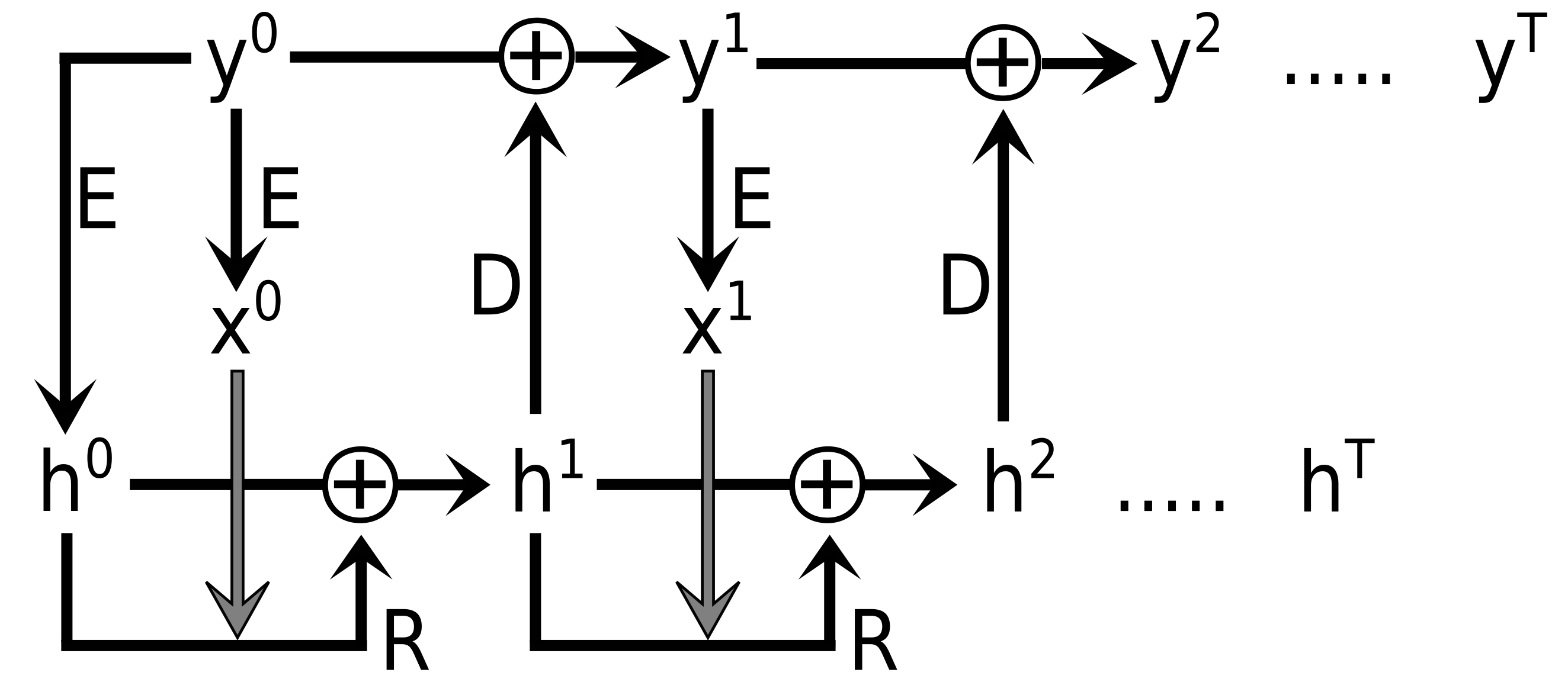}
\end{center}
 \caption{Operations of our weight-tied residual neural network architecture. The neural network modules E, R, and D are learned from data and are shared across individual steps. The initial state $y^0$ is encoded into the initial hidden state $h^0$ and the input $x^0$ of R at time $t=0$. R then predicts additive updates to $h^t$ by using $h^t$ and $x^t$. At each step, D outputs relative updates, which are used to evolve system states $y$ via Euler integration. Sequences $y^t$ and $h^t$ are calculated iteratively, where the number of steps T correlate linearly to simulation time.}
 \label{fig:resnet}
\end{figure}

Our proposed model for prediction of collision outcomes can be interpreted as a residual neural network \citep[ResNet;][]{he16}. ResNets were originally introduced to ease the training of deep neural networks. The "residual" aspect refers to reformulating neural network layers as learning residual functions with reference to the layer inputs, instead of learning unreferenced functions. Comparable to LSTMs without forget gate, ResNets efficiently mitigate the vanishing gradient problem.

We refer to our architecture as RES herein. In contrast to a classical ResNets, learned parameters in our model are shared across different steps (see \cref{fig:resnet}). Our architecture treats temporal dynamics consistently by evolving system states in an autoregressive manner. Individual steps of a trained model can be interpreted as evolving the system for a fixed, but learned time interval. This approach is comparable to explicit iterative methods such as the Euler method or the Runge-Kutta method \citep[][]{runge, kutta}. Ideally, smaller stepsizes should allow for better modeling of physical processes and should thus lead to better performance at the cost of computational resources. Our architecture allows for flexible prediction of system states at different times $T$ by taking the respective number of update steps. Our architecture is auto-regressive, i.e., only requiring the initial state $y^0$ and the number of steps $n_\mathrm{steps} = s_h \times T$ as input. The hyperparameter $s_h$ allows to take different accuracy-speed trade-offs by adjusting the temporal resolution (i.e., the stepsize).

We want to stress that predicted sequences $y^t$ and $h^t$ (with $t\in [0, n_\mathrm{steps}]$) should not be considered as time sequences per se, but may nevertheless be closely related / correlated to time sequences, especially when considering our task of predicting system states at different points in time. We incorporate this close temporal correlation by choosing the number of steps to correlate linearly to the simulated time of the respective SPH simulations. This strong assumption may require additional and more detailed consideration in future work.

Our entire model architecture can be formalized as follows:
\begin{ceqn}
\begin{align}
    (h^t, x^t) &= E(y^t, \phi_{E}) \\
    h^t &= h^{t-1} + R(h^{t-1}, x^{t-1}, \phi_{R}) \\
    y^t &= y^{t-1} + D(h^t, \phi_{D})
\end{align}
\end{ceqn}

$E$, $R$, and $D$ correspond to the encoder module, residual module, and decoder module, respectively. Hidden states that are predicted by E are only used for the initial hidden state $h^0$. Our architecture avoids the vanishing gradient problem via additive updates to hidden states $h$ and physical states $y$. Additive updates to the physical states $y$ can be interpreted as Euler discretization of a continuous transformation and is closely related to the works of \citet{chen18}, \citet{he16}, and \citet{srivastava15}. $E$, $R$, and $D$ are feed-forward neural networks\footnote{historically also referred to as multi-layer perceptrons} \citep[][]{rosenblatt61, ivakhnenko65} with learnable parameters $\phi_{E}$, $\phi_{R}$, and $\phi_{D}$.

\subsubsection{Baseline models}
We choose three baseline methods for comparison with our newly proposed RES model class. Existing work \citep{cambioni19, cambioni21, emsenhuber20, timpe20} use FFNs as regressors for collision outcomes. We thus choose an FFN as our first baseline. We choose a linear regression model (LIN) as our second baseline to study the benefit of deep learning models compared to a simple, data-driven model. The third baseline is perfect inelastic merging (PIM), which is still widely used in astrophysical problems involving collisions because the method is purely analytic and fast. PIM assumes a perfect inelastic collision of target and projectile, always leading to a single surviving body, and conserving mass and momentum of the system by design.

To enable learning non-linear mappings, artificial neural networks require so-called activation functions, which are applied element-wise to individual neurons usually after calculating the matrix-vector product for the respective layers. Due to its sound theoretical advantage compared to other activation functions, we use the scaled exponential linear unit (SELU) activation function \citep{klambauer17} for hidden layers and linear activation functions for output layers of our deep learning models. SELU activations have self-normalizing properties, where neuron activations automatically converge towards zero mean and unit variance in the case of many hidden layers, leading to substantial advantages for training, regularization, and robustness when compared to other approaches. An optional rotation module can be incorporated into the ML models as additional pre -and postprocessing steps, rendering the models rotation-equivariant (see \cref{sec:rotation_equivariance} for details).

Although our SPH results naturally contain approximations and assumptions about the simulated physical processes, and are also subject to typical numerical inaccuracies, we define the SPH data as our ground truth. This definition is generally motivated by the fact that hydrodynamical simulations are currently considered the most accurate method for planetary collision treatment.

\subsubsection{ML experiment setup}
\label{sect:ml_exp_setup}
We split our data into a development set and a test set. The development set includes approximately 88 per cent of the data (8927 datapoints) and consists of training -and validation splits, whereas the test set covers the remaining 12 per cent (1237 datapoints). The entire dataset contains 10164 datapoints. Using the development set, we perform 5-fold crossvalidation\footnote{Folds are non-intersecting, same-sized subsets of a dataset. For 5-fold crossvalidation, a total of 5 models are trained independently. Each training session consists of training on 4 folds, whereas the remaining 5th fold is used for validation.} \citep{hastie17} for all experiments, allowing to calculate confidence intervals for our results. All training and validation splits share the same data distribution. Note that validation data is inappropriate for estimating performance on future data because validation data is used for hyperparameter optimization, which can be a source of information leakage. A holdout test set is required to estimate performance on completely new, unseen data.

Let us recap that our dataset covers the parameter space as defined in \cref{tab:param_space}. Although the parameter space is carefully chosen, parameters of real collisions are naturally not strictly limited to our defined parameter ranges \citep{quionero09}, i.e., so-called out-of-distribution (o.o.d.) datapoints. In practice, ML models often fail to generalize to such o.o.d. datapoints. In order to study o.o.d. generalization of our ML models, we establish an o.o.d. test set. We expect that problem-specific models have better o.o.d. generalization capabilities compared to general-purpose models \citep{mitchell80}. It is widely known that the impact velocity and the impact angle are two of the most important parameters in the context of planetary collisions. Thus, we manually select 4 regions in the impact angle -- impact velocity space that compose our o.o.d. test set (see \cref{tab:test_set} and \cref{fig:post_classes}). We use this o.o.d. test set as our default test set in experiments unless stated otherwise.

\begin{table}
\begin{center}
\begin{tabular}{l|l|l|l|l} 
 parameter & region 1 & region 2 & region 3 & region 4 \\ 
 \hline
 $\alpha_\mathrm{min} [\mathrm{deg}]$ & 10 & 65 & 80 & 0\\
 $\alpha_\mathrm{max} [\mathrm{deg}]$ & 30 & 75 & 90 & 20\\
 $v_\mathrm{imp, min} [v_\mathrm{esc}]$ & 1.5 & 2 & 1 & 6\\
 $v_\mathrm{imp, max} [v_\mathrm{esc}]$ & 2.5 & 4 & 2 & 8
\end{tabular}
\end{center}
\caption{Selected regions for the out-of-distribution (o.o.d.) test set. We select regions that may result in qualitatively different outcomes compared to the development set. The test set contains about 12 per cent of all datapoints.}
\label{tab:test_set}
\end{table}

Our dataset $D$ consists of $N=10164$ tuples $(y^0_i, z^T_i), \ \ i \in [1, N]$, representing initial system states (at $t=0$) and final system states (at $t=T$). For our supervised learning task, $y^0_i$ and $T$ are used as model inputs whereas $z^T_i$ are used as ground truth labels. For intermediate states $0<t<T$ holds. $D$ is split into a development set $D_\mathrm{dev}$ and a test set $D_\mathrm{test}$. Our training and validation splits are derived from $D_\mathrm{dev}$ depending on the respective crossvalidation fold.
\begin{ceqn}
\begin{align}
&D=\{(y^0_1, z^T_1), (y^0_2, z^T_2), ..., (y^0_N, z^T_N)\} \\
&D_\mathrm{dev} \subset D,\ \ D_\mathrm{test} \subset D, \ \ D_\mathrm{dev} \cap D_\mathrm{test} = \{\}
\end{align}
\end{ceqn}

We perform data preprocessing to transform features into appropriate value ranges for ML. We apply feature-wise normalization $x_\mathrm{ML} = \frac{(x_\mathrm{phys} - \mu)}{\sigma}$ to transform data $x_\mathrm{phys}$ given in SI units into data $x_\mathrm{ML}$, whose value ranges are better suited for ML. We set $\mu=0$ for all features. Note that the barycentre of the system still remains at the origin of the coordinate system for all datapoints even after normalization. \cref{tab:mode_features} summarizes our ML features, along with normalization hyperparameters $\sigma$. Note that $\mu$ and $\sigma$ implicitly define the importance between different sub-tasks during ML model training. Importance of sub-tasks can be further adjusted via introducing dedicated weights for the corresponding loss terms. The detailed preprocessing pipeline can be found in the provided source code (see \cref{data_avail}).

Note that although accurate tracking of the rotation state is important for many aspects of planet formation and evolution modeling, we do not include rotation in our model predictions. This is because it is non-trivial to derive physically reliable (and unique) post-collision rotation states within our post-processing chain for SPH collision simulations. A major point to consider is that our definition of a remnant is not restricted to a single physically connected fragment, but also includes all gravitationally bound fragments in addition. In reality, these fragments may or may not be actually accreted at some later point in time, or interact otherwise with each other. Nevertheless, we consider including these fragments into the definition of \textit{remnant} as the best possible option, considering the alternative of simply ignoring them. This comes on top of the general issue that approximate rotational equilibrium has to be achieved after the collision in order to extract a reliable rotation state, which is highly scenario-dependent in terms of the relevant dynamics and timescales. Considering those difficulties, we decided not to include the rotation state in our ML model predictions. Therefore, while pre-collision rotation is fully accounted for, we do not attempt to predict post-collision rotation states in this work.

\begin{table*}
\begin{center}
\begin{tabular}{ l|l|l|l|l|l } 
 state & feature & dim & description & $\sigma$ (ours) & $\sigma$ (Timpe)\\ 
 \hline
 initial & $N_\mathrm{tot} [1]$ & 1 & number of SPH particles & 5e+4  & 2.3e+5\\
 & $M_\mathrm{tot} [kg]$ & 1 & total mass & 1e+25  & 100\\
 & $\gamma [1.]$ & 1 & mass ratio $m_p/m_t$ & 1  & 1\\
 & $\zeta_p [1.]$ & 2 & material fractions projectile & 1 & 1\\
 & $\zeta_t [1.]$ & 2 & material fractions target & 1 & 1\\
 & $rot_p [rad/s]$ & 3 & rotation axis projectile & 6.5e-05  & 100\\
 & $rot_t [rad/s]$ & 3 & rotation axis target & 6.5e-05  & 100\\
 & $x_p [m]$ & 3 & barycentre position projectile & [5e+07, 2e+08, 2e+07]  & [6, 41, 3]\\
 & $v_p [m/s]$ & 3 & barycentre velocity projectile & [2e+03, 1e+04, 6e+02]  & [2, 16, 0.5] \\
 & $x_t [m]$ & 3 & barycentre position target & [5e+07, 2e+08, 2e+07]  & [6, 41, 3]\\
 & $v_t [m/s]$ & 3 & barycentre velocity target & [2e+03, 1e+04, 6e+02]  & [2, 16, 0.5]\\
 \hline
 final & $m_1 [kg]$ & 1 & mass largest remnant & 1e+25 & 100\\
 & $m_2 [kg]$ & 1 & mass 2nd-largest remnant & 1e+25 & 100\\
 & $m_r [kg]$ & 1 & mass rest & 1e+25 & 100\\
 & $\zeta_1 [1.]$ & 2 & material fractions largest remnant & 1 & 1\\
 & $\zeta_2 [1.]$ & 2 & material fractions 2nd-largest remnant & 1 & 1\\
 & $\zeta_r [1.]$ & 2 & material fractions rest & 1 & 1\\
 & $x_1 [m]$ & 3 & barycentre position largest remnant & [5e+07, 2e+08, 2e+07]  & [6, 41, 3]\\
 & $v_1 [m/s]$ & 3 & barycentre velocity largest remnant & [2e+03, 1e+04, 6e+02]  & [2, 16, 0.5]\\
 & $x_2 [m]$ & 3 & barycentre position 2nd-largest remnant & [5e+07, 2e+08, 2e+07]  & [6, 41, 3]\\
 & $v_2 [m/s]$ & 3 & barycentre velocity 2nd-largest remnant & [2e+03, 1e+04, 6e+02]  & [2, 16, 0.5]\\
 & $x_r [m]$ & 3 & barycentre position rest & [5e+07, 2e+08, 2e+07]  & [6, 41, 3]\\
 & $v_r [m/s]$ & 3 & barycentre velocity rest & [2e+03, 1e+04, 6e+02]  & [2, 16, 0.5]
\end{tabular}
\end{center}
\caption{Non-redundant ML features and normalization hyperparameters for feature normalization. Units indicate different physical quantities. All data $x_\mathrm{phys}$ is normalized during preprocessing. Note that since material fractions sum up to 1, only core (iron) and shell (water) fractions are required. Initial rotation speeds of the colliding bodies are encoded via the norms of their respective rotation axes.}
\label{tab:mode_features}
\end{table*}

Our optimization objective is to minimize the mean absolute error (MAE) between model predictions $y^T_i$ and ground truth labels $z^T_i$ over our training data:

\begin{ceqn}
\begin{align}
y^T_i &= f(y^0_i, T, \phi) \\
\mathcal{L} &= \frac{1}{M} \sum_{i=1}^{M}{\xi \cdot ||y^T_i-z^T_i||}
\end{align}
\end{ceqn}

$f: \mathbb{R}^d \to\mathbb{R}^k$ refers to an ML model that regresses final states when given initial states. In this work we focus on handling macroscopic system states for both model inputs and outputs, resulting in $d=25$ and $k=27$. $f$ has learnable parameters $\phi$ that we aim to optimize. $M$ refers to the size of the training split for individual cossvalidation folds.

We treat every unit of mass (i.e., every kilogram) as equally important in our ML task. We account for this treatment by calculating mass-dependent weights $\xi$ in order to re-weight errors for output features that correspond to the largest remnant, 2nd-largest remnant, and the rest (of material), respectively. 

\begin{ceqn}
\begin{align}
\xi &= [\xi_\mathrm{lr},\ \xi_\mathrm{2lr},\ \xi_\mathrm{rest}] \\
\xi &= \left[\frac{m_\mathrm{lr}}{m_\mathrm{tot}}, \frac{m_\mathrm{2lr}}{m_\mathrm{tot}}, \frac{m_\mathrm{rest}}{m_\mathrm{tot}}\right] \\
m_\mathrm{tot} &= m_\mathrm{lr} + m_\mathrm{2lr} + m_\mathrm{rest}
\end{align}
\end{ceqn}

We consider this re-weighting to be essential to accurately reflect the prediction problem, especially if $m_\mathrm{2lr} << m_\mathrm{lr}$ or $m_\mathrm{rest} << m_\mathrm{lr}$. Early experiments without re-weighting lead to poor prediction performance originating from small objects (remnants and fragments). In general, small objects are more difficult to predict compared to large objects. Moreover, the assignment of the 2nd-largest remnant tends to jump in the presence of many small objects, making it almost impossible to predict robustly. This labelling noise then leads to large error gradients, hampering learning significantly. This problem is ameliorated with our re-weighting approach. 

We use identical training hyperparameters for our deep learning models (FFN and RES). Training is performed via stochastic gradient descend \citep{robbins51}, utilizing the backpropagation algorithm \citep{kelley60, rumelhart86}. We use the adamax optimizer \citep{kingma14} with default hyperparameters, a minibatch size of $bs=128$, a constant learning rate of $\eta=0.0005$, and a weight decay of $wd=0.0001$. We apply gradient-norm clipping \citep{pascanu13}, allowing for maximum gradient norms of $n_\mathrm{grad}=50$. Moreover, we use exponential moving average models \citep{polyak90, ruppert88, tarvainen17} with a rate of $r_\mathrm{ema}=0.999$ for validation and testing. We find that the mean squared error (MSE) leads to worse validation performance than the mean absolute error (MAE), which is more robust to outliers. To alleviate the exploding gradient problem as described by \citet{metz21}, we additionally penalize too large activations of hidden states $h_t$ in our RES model. We train each of our models for 5000 epochs, which is sufficient to reach convergence.

Since different ML architectures are inherently difficult to compare, we try to find the best architectures and respective models in terms of validation performance for each model class (i.e., FFN, RES) separately. We optimize hyperparameters manually in an iterative manner (i.e. always optimizing one hyperparameter at a time while keeping others fixed, and repeating the procedure until convergence in validation performance) and dedicate approximately the same amount of time and computational resources to optimize each set of hyperparameters for FFN and RES. \cref{tab:hyperparams} summarizes all optimized hyperparameters for FFN and RES, while LIN has no model-class hyperparameters. In order to prevent information leakage and misleading test performance, we solely perform hyperparameter finetuning based on validation performance. Test performance is measured after model development was completed. In principle, RES allow using intermediate states as additional learning signals. Unless stated otherwise, we only use final states for training to ensure a fair comparison with our baselines.

We use the root mean squared error (RMSE) as our validation metric. For certain applications prediction speed may play a significant role. We note that the RMSE metric does not account for this aspect and thus purely focuses on prediction accuracy. All performance results reported below are obtained by first taking the best RMSE (minimum) over all epochs for every fold individually, then averaging them over the folds. Errors indicate the minimum and maximum (over folds) of best RMSE values. The same procedure holds for classification accuracies, except for first taking the maximum instead of the minimum. Measuring and interpreting RMSE in the data-space is unintuitive in our multi-task setting due to vastly different value ranges of individual quantities. Thus, RMSE is measured in ML feature-space. Moreover, since RMSE values have to be interpreted in consideration of the overall ML task, we recommend comparing reported results relative to each other.

We use the balanced accuracy score for validating classification performances of our models. Balanced accuracy in the multi-class classification setting is defined by taking the average of true positive rates for individual classes. The true positive rate is also referred to as sensitivity or recall. Considering the strong class-imbalances that are typically present in planetary collision data, we consider the balanced accuracy score to be much more problem-focused and more applicable compared to the unbalanced accuracy score.

\subsubsection{Efficiency considerations of ML}
\label{sec:efficiency_considerations}
In practice, researchers are interested at which point using ML starts to pay off compared to classical approaches for a fixed computational budget. Consider the goal of predicting $m$ collision outcomes. For classical approaches such as PIM or direct SPH simulations we only need to consider inference times, which scale linearly with $m$. On the other hand, ML requires consideration of data generation, training, and inference. In general, the required computation times for these three components are mostly independent from each other. In our case, data generation requires by far the most time, followed by training. Finally, ML inference requires only a tiny fraction of the overall computation budget. Thus, we recommend using ML approaches in case of extensive inference, i.e., large $m$. Let us define the three computation times $\tau_\mathrm{d}$ for generating one datapoint, $\tau_\mathrm{t}$ for ML model training, and $\tau_\mathrm{i}$ for inference of one datapoint. In our case, we consider $\tau_\mathrm{t}$ as the total wall-clock training time for 5 folds, each having 5000 epochs. $N$ refers to the training dataset size. We can calculate for which $m$ the use of ML pays off (i.e., $T_\mathrm{ML} < T_\mathrm{CL}$) when comparing the overall computation times $T_\mathrm{CL}$ for classical approaches with computation times $T_\mathrm{ML}$ for ML approaches:

\begin{ceqn}
\begin{align}
    T_\mathrm{CL} &= m \times \tau_\mathrm{i, CL} \\
    T_\mathrm{ML} &= N \times \tau_\mathrm{d} + \tau_\mathrm{t} + m \times \tau_\mathrm{i, ML} \\
    m &> \frac{N \times \tau_\mathrm{d} + \tau_\mathrm{t}}{\tau_\mathrm{i, CL} - \tau_\mathrm{i, ML}}
\end{align}
\end{ceqn}

\section{Results}
\subsection{SPH collision data}
The provided SPH data serves as the basis for ML models in order to solve the collision treatment problem accurately and fast. To our knowledge, this dataset is the first of its kind to combine different aspects such as object rotation, realistic object models including water layers, and providing time-series data. All data can be freely accessed (see \cref{data_avail}). 

\begin{figure}
 \includegraphics[width=\columnwidth]{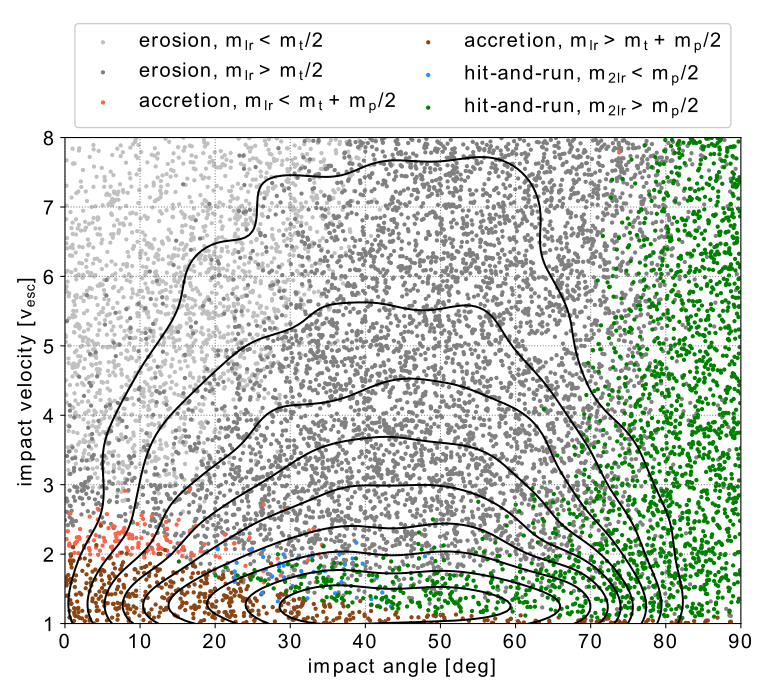}
 \caption{Overview of collision outcomes in impact angle -- impact velocity space. Each datapoint represents a simulation in our SPH dataset. Colors indicate major outcome regimes: erosion (grayish), accretion (reddish), and hit-and-run (greenish). Each regime is further divided into sub-categories, depending on remnant masses. The contour overlay indicates collision statistics in a realistic dynamical environment, obtained from $N$-body simulations by \citet{burger20}, see \cref{sect:nbody}. Contour levels correspond to iso-proportions of the density (in 10 per cent steps).}
 \label{fig:stewart}
\end{figure}

Our results are consistent with \citet{leinhardt12} and \citet{stewart12} in identifying three major outcome regimes, erosion, accretion, and hit-and-run. We define these regimes as
\begin{itemize}
\item erosion:\quad $m_{lr} < m_{t}$
\item accretion:\quad $m_{lr} > m_{t} \land m_{2lr} \le 0.1 m_{p}$
\item hit-and-run:\quad $m_{lr} > m_{t} \land m_{2lr} > 0.1 m_{p}$
\end{itemize}
where subscripts indicate the largest remnant, 2nd-largest remnant, target, and projectile, respectively. Each of these regimes can be further divided into sub-classes, based on thresholds for remnant masses, as defined in \cref{tab:classes}, and plotted in \cref{fig:stewart}. Erosion typically results from high impact velocities and/or low impact angles, whereas accretion mostly emerges for lower impact velocities. Hit-and-run either results from high impact angles, or from a combination of lower impact angles, large-enough projectile-to-target mass ratios, and impact velocities that are low enough to avoid global disruption but high enough to avoid an accretion-type outcome.


\begin{table}
\begin{center}
\begin{tabular}{l|l|l|l} 
 class & sub-class & random & realistic \\ 
 \hline
 erosion & $m_{lr}<m_t/2$ & 18.3\% (1856) & 3.0\% (151) \\
 erosion & $m_{lr}>m_t/2$ & 55.1\% (5600) & 58.5\% (2968) \\
 accretion & $m_{lr}<m_t+m_p/2$ & 1.3\% (135) & 2.0\% (102) \\
 accretion & $m_{lr}>m_t+m_p/2$ & 5.2\% (526) & 5.5\% (281) \\
 hit-and-run & $m_{2lr}<m_p/2$ & 0.4\% (40) & 4.4\% (221) \\
 hit-and-run & $m_{2lr}>m_p/2$ & 19.7\% (2007) & 26.7\% (1353)
\end{tabular}
\end{center}
\caption{Class counts for different outcome regimes, and for random and realistic collision parameters. The three major outcome regimes are each further divided into sub-classes, based on remnant masses. Random parameters are obtained from uniform, random sampling, whereas realistic conditions are obtained from dynamically consistent $N$-body simulations.}
\label{tab:classes}
\end{table}

\subsection{ML experiments}
Below we present our results for the experiments described in \cref{sect:ml_exp_setup}. Results indicated by $^*$ are statistically significant ($p<0.05$) according to a Wilcoxon test when comparing FFN with RES for the respective experiments.
\subsubsection{Performance}

\begin{table*} 
\begin{center}
\begin{tabular}{ r|r|r|r|r|r|r|r } 
 method & notes & mass & material & position & velocity & total & accuracy \\[0.5ex] 
 \hline
 PIM & o.o.d. test & $0.1501^{+0.0000}_{+0.0000}$& $0.0605^{+0.0000}_{+0.0000}$& $0.3046^{+0.0000}_{+0.0000}$& $0.2229^{+0.0000}_{+0.0000}$& $0.2450^{+0.0000}_{+0.0000}$& $0.1667^{+0.0000}_{+0.0000}$\\[0.5ex]
 LIN & o.o.d. test & $0.0460^{+0.0009}_{-0.0012}$& $0.0250^{+0.0002}_{-0.0002}$& $0.1880^{+0.0007}_{-0.0007}$& $0.1509^{+0.0015}_{-0.0014}$& $0.1479^{+0.0009}_{-0.0010}$& $0.2340^{+0.0043}_{-0.0057}$\\[0.5ex]
 FFN & o.o.d. test & $0.0121^{+0.0008}_{-0.0008}$& $0.0129^{+0.0006}_{-0.0003}$& $0.0487^{+0.0014}_{-0.0012}$& $0.0433^{+0.0010}_{-0.0010}$& $0.0408^{+0.0008}_{-0.0009}$& \bm{$0.4964^{+0.0480}_{-0.0538}$}\\[0.5ex]
 RES & o.o.d. test & \bm{$^*0.0108^{+0.0003}_{-0.0005}$}& \bm{$0.0127^{+0.0002}_{-0.0002}$}& \bm{$^*0.0386^{+0.0015}_{-0.0018}$}& \bm{$0.0428^{+0.0004}_{-0.0003}$}& \bm{$^*0.0364^{+0.0009}_{-0.0010}$}& $0.4887^{+0.0160}_{-0.0139}$\\[0.5ex]
 \hline
 FFN & o.o.d. test, +labels & $0.0122^{+0.0009}_{-0.0010}$& \bm{$0.0136^{+0.0002}_{-0.0004}$}& $0.0500^{+0.0015}_{-0.0020}$& $0.0433^{+0.0005}_{-0.0004}$& $0.0416^{+0.0009}_{-0.0010}$& $0.5309^{+0.0673}_{-0.1678}$\\[0.5ex]
 RES & o.o.d. test, +labels & \bm{$^*0.0107^{+0.0008}_{-0.0009}$}& $0.0136^{+0.0002}_{-0.0002}$& \bm{$^*0.0372^{+0.0012}_{-0.0014}$}& \bm{$0.0426^{+0.0011}_{-0.0006}$}& \bm{$^*0.0358^{+0.0007}_{-0.0009}$}& \bm{$0.5311^{+0.0548}_{-0.0311}$}\\[0.5ex]
 \hline
 FFN & o.o.d. test, single & \bm{$0.0083^{+0.0003}_{-0.0004}$} & $0.0098^{+0.0002}_{-0.0002}$& $0.0422^{+0.0009}_{-0.0010}$ & \bm{$0.0409^{+0.0007}_{-0.0008}$}& - & $0.4720^{+0.0417}_{-0.0470}$\\[0.5ex]
 RES & o.o.d. test, single & $0.0083^{+0.0008}_{-0.0004}$ & \bm{$^*0.0090^{+0.0003}_{-0.0003}$} & \bm{$^*0.0364^{+0.0024}_{-0.0017}$} & $0.0422^{+0.0004}_{-0.0010}$& - & \bm{$^*0.5165^{+0.0349}_{-0.0228}$}\\[0.5ex]
 \hline
 PIM & i.i.d. test, Timpe & $0.2088^{+0.0000}_{+0.0000}$& $0.1925^{+0.0000}_{+0.0000}$& $0.3691^{+0.0000}_{+0.0000}$& $0.2880^{+0.0000}_{+0.0000}$& - &$0.1667^{+0.0000}_{+0.0000}$\\[0.5ex]
 LIN & i.i.d. test, Timpe & $0.0518^{+0.0003}_{-0.0002}$& $0.0473^{+0.0003}_{-0.0005}$& $0.2105^{+0.0005}_{-0.0002}$& $0.1693^{+0.0002}_{-0.0005}$& - &$0.3160^{+0.0025}_{-0.0015}$\\[0.5ex]
 FFN & i.i.d. test, Timpe & $0.0132^{+0.0002}_{-0.0001}$& $0.0144^{+0.0002}_{-0.0001}$& $0.0877^{+0.0017}_{-0.0012}$& $0.0631^{+0.0006}_{-0.0004}$& - &$0.4675^{+0.0039}_{-0.0021}$\\[0.5ex]
 RES & i.i.d. test, Timpe & \bm{$^*0.0126^{+0.0003}_{-0.0004}$}& \bm{$0.0141^{+0.0006}_{-0.0006}$}& \bm{$^*0.0840^{+0.0037}_{-0.0036}$}& \bm{$0.0628^{+0.0007}_{-0.0009}$}& - &\bm{$0.4690^{+0.0026}_{-0.0023}$}
\end{tabular}
\end{center}
\caption{Test performance (RMSE and balanced accuracy, both measured on the final state) of different approaches for planetary collision treatment. Classification is performed as a postprocessing step on top of predicted masses. We assume the SPH simulation data to be the ground truth. For single-task learning (last 6 rows), each entry corresponds to the performance of individually trained ML models and column headers indicate optimized tasks respectively. We use data from \citet{timpe20} to obtain the results in the last 4 rows. Best results are indicated in bold, whereas * indicate statistically significant results according to a Wilcoxon test (comparing FFN and RES). Our proposed RES model outperforms the other baseline methods in most experiments.}
\label{tab:exp_main}
\end{table*}

We compare commonly used methods for planetary collision treatment with our proposed RES model and summarize the results in \cref{tab:exp_main} and \cref{tab:train_and_vali_performance}. Our o.o.d. test set consists of datapoints within manually selected regions in the impact angle -- impact velocity space (dashed regions in \cref{fig:post_classes}).

\begin{figure}
 \includegraphics[width=\columnwidth]{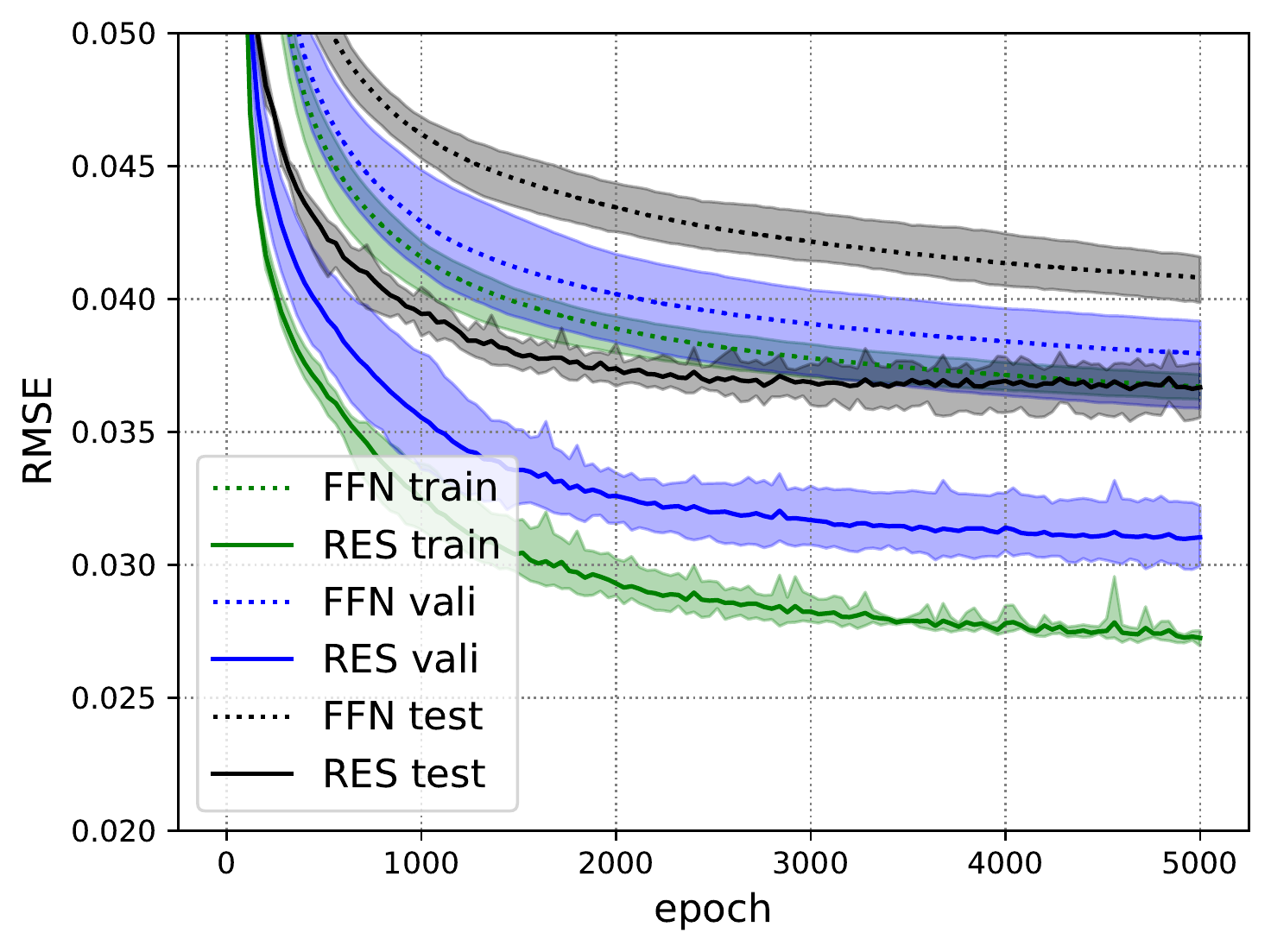}
 \caption{Learning curves depicting the training, validation, and test performance for our multi-task objective. RMSE is measured on the final state ($t=T$). Shaded regions indicate the minimum and maximum performance over the 5 crossvalidation folds. Performance increases significantly during the first 1000 epochs before converging. RES significantly outperforms the FFN baseline.}
 \label{fig:training_curves}
\end{figure}

All deep learning models outperform the PIM and LIN baselines by a large margin. Improved performance over PIM was expected, since it is an analytic model that applies very simplistic assumptions on collision dynamics. However, we still regard PIM as a useful method in case of limited computational resources. Improved results over the LIN baseline were also expected since it assumes that the data is linearly dependent. RES perform best amongst the deep learning approaches that we studied, significantly outperforming the FFN baseline, also illustrated in \cref{fig:training_curves}. In general, our deep learning models generalize well to the o.o.d. test set, indicating that they might even generalize beyond our covered parameter space (\cref{tab:param_space}). RES consistently outperforms the FFN baseline in terms of RMSE on the o.o.d. test set.

We do not observe performance gains when using intermediate states as additional labels during training. We observe a shrinkage effect (regression to the dataset mean) in model predictions when using intermediate states, which may harm performance measured on the final state. Moreover, non-converged intermediate states may have a relatively high labelling noise due to the discretization into remnants, potentially making intermediate states more difficult to predict compared to final states. We also believe that intermediate macro states are somewhat redundant, unless operating on a microscopic scale, i.e., directly learning from SPH particle representations or similar.


Direct prediction of final states with FFNs only allows analysis of hidden representations, which are typically too abstract to interpret due to their high-dimensional, learned nature. In addition to the analysis of hidden representations, our RES architecture allows analyzing predicted intermediate states, opening up an entry point for interpreting model predictions in more detail. This is also illustrated in \cref{fig:trajectories}. Although our learning objective contains no incentive for predicted remnant trajectories to align with ground truth trajectories, we observe spatio-temporally continuous transitions from initial to final states. This indicates that steps in RES may correlate to the temporal evolution of the physical system to a certain extent, even though our results do not allow for strong conclusions.

We also train our ML models on the planetary collision dataset provided by \citet{timpe20} (single-task setting) and report performance results on the independent and identically distributed (i.i.d.) test set in \cref{tab:exp_main}. Without performing any additional hyperarameter finetuning, we observe that RES outperforms the other baseline methods in all sub-tasks, verifying that our RES model is dataset-agnostic. However, our results are not statistically significant for 2/4 sub-tasks.

\subsubsection{Efficiency}

\begin{table}
\begin{center}
\begin{tabular}{ r|r|r|r|r|r } 
 method & \#param & \#hyper & $\tau_\mathrm{d}$ & $\tau_\mathrm{t}$ & $\tau_\mathrm{i}$ \\ 
 \hline
 PIM & 0 & 0 & - & - & <1 s\\
 LIN & 729 & 0 & 0.75 h & <1 s & <1 s\\
 FFN & 22203 & 2 & 0.75 h & 11.95 h & <1 s\\
 RES, $s_h=1$ & 64417 & 3 & 0.75 h & 33.51 h & <1 s\\
 RES, $s_h=2$ & 64417 & 3 & 0.75 h & 44.78 h & <1 s\\
 RES, $s_h=3$ & 64417 & 3 & 0.75 h & 56.45 h & <1 s\\
 RES, $s_h=4$ & 64417 & 3 & 0.75 h & 66.74 h & <1 s\\
 SPH & 0 & $\sim$5 & - & - & 0.75 h
\end{tabular}
\end{center}
\caption{Number of learnable parameters and number of optimized hyperparameters of our methods, as well as typical computation times for data generation, training, and inference on the same hardware (GPU: Nvidia GTX 1080Ti). $\tau_d$ and $\tau_i$ are the average computation times for generating one datapoint and model inference for one datapoint, whereas $\tau_t$ is the total wall-clock training time for 5 folds, each having 5000 epochs. The required time for data generation is $\tau_D = N \times \tau_d$ and takes the largest part of the overall computational budget. $s_h$ is the number of RES model steps taken per simulated hour.}
\label{tab:time_and_s_h_ablation}
\end{table}

We report the required computational resources of our methods in \cref{tab:time_and_s_h_ablation}. As expected, PIM marks one extremum of the accuracy-speed trade-off, requiring the least amount of computational resources. On the other hand, SPH marks the other extremum, scaling badly with $m$ (see \cref{sec:efficiency_considerations}). ML models cover intermediate trade-off regions, depending on the model class and hyperparameter choices. ML models are approximately 4 magnitudes faster in inference compared to SPH (both running on a single GPU), allowing ML models to be efficiently used in large-scale planetary evolution simulations. Considering the results from \cref{tab:time_and_s_h_ablation}, for our deep learning models it holds that $N \times \tau_{d} \gg \tau_t \gg N \times \tau_i$ for $N=10164$. This indicates that the most effective way to save computation when using ML is by requiring less training data, i.e., small $N$. However, using less data will inevitably lead to worse performance in terms of RMSE and to degrading generalization. 

Therefore, we perform ablation studies using different training dataset sizes to investigate data-efficiency of our deep learning models and report the corresponding o.o.d. test performances in \cref{tab:data_ablation}. The results indicate that RES requires 50 per cent less data to achieve comparable performance to the FFN baseline. In other words, using RES is much more efficient than the FFN baseline in the case of comparable performance. Once trained, ML methods are practically as fast as PIM, while maintaining high prediction accuracy.

\begin{table}
\begin{center}
\begin{tabular}{ r|r|r|r } 
 method & 100\% data & 50\% data & 25\% data \\[0.5ex] 
 \hline
 FFN & $0.0408^{+0.0008}_{-0.0009}$ & $0.0442^{+0.0004}_{-0.0005}$ & $0.0516^{+0.0012}_{-0.0012}$\\[0.5ex]
 RES & \bm{$^*0.0364^{+0.0009}_{-0.0010}$} & \bm{$^*0.0398^{+0.0009}_{-0.0013}$} & \bm{$^*0.0469^{+0.0013}_{-0.0010}$}
\end{tabular}
\end{center}
\caption{RMSE of deep learning models on the fixed-size o.o.d. test set when using 100, 50, and 25 per cent of training data. RES requires 50 per cent less data to achieve similar performance compared to FFN.}
\label{tab:data_ablation}
\end{table}

\Cref{tab:time_and_s_h_ablation} also summarizes the number of learnable parameters and the number of model class hyperparameters of all collision treatment methods we studied in this work. ML training requires additional hyperparameters such as $bs$, $\eta$, $wd$, and $n_\mathrm{grad}$. The number of learnable parameters for LIN is fully determined by the ML task, i.e., the dimensions of input and label vectors. For deep learning methods, the number of learnable parameters depends on the hyperparameter choices that ultimately define model architectures. We optimize hyperparameters w.r.t. validation performance and perform extensive ablation studies for both FFN and RES to verify their optimal hyperparameters. We consider the reported model capacities (i.e. the number of learnable parameters) to be optimal in terms of validation performance for our data. We find that the optimal FFN architecture requires less learnable parameters compared to the optimal RES architecture. In particular, increasing FFN's size does not improve its prediction performance anymore. We refer to \cref{sec:hyperparams} for more details about hyperparameters. The number of hyperparameters for the direct-SPH method accounts for the most important method-specific aspects such as the smoothing length and settings related to the equation of state of the simulated material.

We study the effects of multi-task learning and single-task learning on model performance and report ablation studies in \cref{tab:exp_main}. Multi-task learning is computationally more efficient by design (i.e., prediction speed and required parameters), requiring only a single model for predicting several different modalities. However, we observe a performance decrease when comparing multi-task learning (1 model with 4 tasks) to single-task learning (4 models, each with 1 task) in favor of single-task learning for both FFN and RES. We perform statistical significance tests (Wilcoxon) between single-task and multi-task experiments. We find single-task significantly ($p<0.05$) outperforms multi-task for almost all sub-tasks for both FFN and RES. Two exceptions are performances for the position and velocity tasks on the o.o.d. test set for the RES model. We conclude that our hypothesis of improved generalization due to mutual benefit via exploiting shared representations does not hold in our experiments. Since we optimized model hyperparameters for the multi-task setting, single-task models might be even better with further finetuning. RES outperforms FFN in 4/4 sub-tasks for multi-task learning, but only in 2/4 for single-task learning, indicating that RES benefits a bit more from multi-task learning.

\begin{figure}
 \includegraphics[width=\columnwidth]{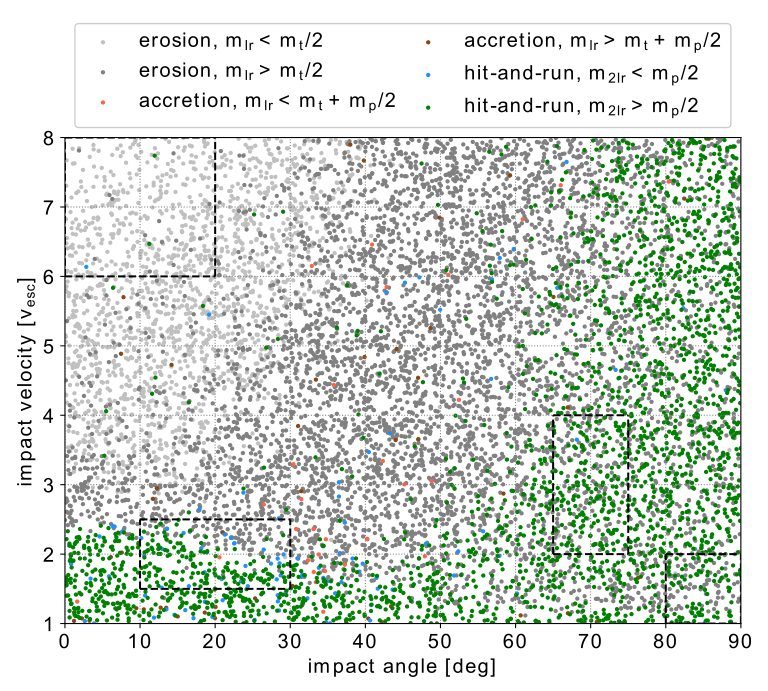}
 \caption{Classification results on validation and o.o.d. test data using predicted remnant masses of the RES model. Datapoints for validation are obtained by combining the respective validation splits that originate from different crossvalidation folds, whereas o.o.d. test set regions are marked with dashed boxes. The model learned to differentiate between typical collision scenarios~/~outcomes and generalizes to the o.o.d. test set.}
 \label{fig:post_classes}
\end{figure}

We verify that our regression-approach is suited to perform classification of different collision scenarios as a postprocessing step, avoiding the need for two-step classification-regression approaches. Classification accuracies are calculated on top of regression results w.r.t. the 6 different classes as defined in \cref{tab:classes}. Balanced accuracies are reported in \cref{tab:exp_main}. \Cref{fig:post_classes} visualizes predicted collision outcomes based on predicted masses of the largest and 2nd-largest remnants.

We observe that our ML models typically tend to mispredict actual accretion scenarios as hit-and-run, ultimately resulting in misclassifications in postprocessing. This is pronounced for low-velocity (close to $v_\mathrm{esc}$) collisions, and particularly for lower impact angles ($\lesssim 30^\circ$), and directly visible when comparing \cref{fig:stewart} and \cref{fig:post_classes}. We assume that this mistreatment stems from the relative under-representation of accretion scenarios in our data (see \cref{tab:classes}), which results in models having poor classification performance for the respective scenarios. This is a common problem in imbalanced classification tasks and can be tackled with different approaches such as generating more data for under-represented classes, oversampling of under-represented classes, regularizing the model during training, or introducing problem-specific model architectures. Note that the balanced accuracy score succeeds in reflecting the reduced classification performance on under-represented classes. In contrast, the unbalanced accuracy score typically tends to be much higher, since under-represented classes are not accounted for properly.

\section{Conclusion}
We perform $N$-body and SPH simulations to tackle the problem of accurate and fast treatment of planetary collisions with ML methods. We use the SPH data to employ a simple but problem adapted ML model for predicting masses, material fractions, positions, and velocities of the two largest post-collision remnants and the remaining debris. Our model helps modeling of temporal dynamics by evolving system states in an autoregressive manner, which closely resembles the data-generating process. The model allows for flexible prediction of post-collision states at different times and can be employed for collision treatment within existing $N$-body simulation frameworks.

We summarize our experiment results by comparing the performance of our two best methods, the FFN baseline and our proposed RES model, for all experiments on the respective test sets. We count a total of 24 comparisons. RES outperforms FFN in 20/24 cases. In 13/20 cases for RES and 0/4 cases for FFN, improvements are statistically significant ($p<0.05$). Moreover, RES is also more data-efficient, achieving similar performance while requiring approximately half as much data compared to FFN. Although multi-task learning is more computationally efficient than single-task learning, we do not observe improved generalization induced by shared representations.

We demonstrate that the FFN baseline is outperformed by our RES model due to its problem-adapted algorithmic bias (i.e. its autoregressive structure), which is better suited for modeling the underlying physical processes in planetary collisions over time (see also Sect.~\ref{sec:autoregressive_models}). The optimal RES architecture requires more learnable parameters than the optimal FFN architecture because RES consists of three neural network modules, whereas FFN consists of only one module. However, we find that increasing FFN's size does not increase its performance anymore. This finding indicates that the performance improvement from RES originates from its better algorithmic bias rather than from its relatively larger number of learnable parameters. Moreover, the superiority of RES is apparent in both our own data and data from \citet{timpe20}, indicating a general trend rather than a data-specific effect.

Measuring the actual effects of systematic errors induced by ML model predictions still remains an open topic in the context of planetary formation and evolution modeling. Thus, a natural follow-up work could be to test various ML methods for collision treatment in such simulations.

Beyond studying the basic task of collision outcome prediction, our data and methods also open up further interesting lines of research related to planet formation in general. This includes studying inverse problems or focusing on specific collision scenarios \citep{canup12, chau18}. Other possible directions are the extension of our methods to different regimes such as small bodies or objects including (proto-)atmospheres, probably requiring to extend the underlying physics model. The latter can include more sophisticated equations of state for more realistic thermodynamics and advanced models for material strength to accurately simulate solid-body behavior.

ML models might benefit from learning directly based on microscopic representations, i.e., fragments or even down to the SPH particle level, and thereby improve aspects regarding generalization and interpretability. Incorporating certain aspects like symmetries, conserved quantities, and sophisticated numerical approaches that have been developed in recent years \citep{alet21, satorras21, brandstetter22, hoedt21, chen20} could be promising directions for further improvements of ML architectures. For example, graph neural networks \citep[GNNs;][]{scarselli09, defferrard16, kipf17} and regression forests \citep{ho95} have been successfully applied to the approximation of numerical simulations \citep{gonzalez20, martinkus20, pfaff20, mayr21, ladicky15}. Efficient ML approaches begin to replace traditional PDE solvers in the context of hydrodynamic simulations \citep{li20fourier, li20neural}. Residual neural networks \citep{he16} showed promising results in modelling complex dynamical processes by formulating the neural network layer-structure as a continuous-depth model \citep{queiruga20} in the context of neural ODEs \citep{chen18, kidger22}.  ML methods allow accurate collision modeling at scale, while at the same time being orders of magnitude faster compared to classical, non data-driven approaches.

\section*{Acknowledgements}
C.B.\ and C.M.S.\ appreciate support by the German Research Foundation (DFG) project 398488521. C.B.\ acknowledges support by the Austrian Science Fund (FWF) project S11603-N16. T.I.M.\ acknowledges support from the Austrian Science Fund (FWF): P33351-N. The authors acknowledge support by the High Performance and Cloud Computing Group at the Zentrum für Datenverarbeitung of the University of Tübingen, the state of Baden-Württemberg through bwHPC and the German Research Foundation (DFG) through grant no. INST 37/935-1 FUGG.

We thank Sepp Hochreiter, Günter Klambauer, Johannes Brandstetter, Pieter-Jan Hoedt, Max Zimmermann, Rudolf Dvorak, Kajetan Schweighofer, Andreas Mayr, Miles Timpe, and Christian Reinhardt for contributing valuable feedback and comments. Simulations for our datasets made use of the REBOUND code \citep{2012A&A...537A.128R}. We thank the developers of Python \citep{python}, Pytorch \citep{pytorch}, Matplotlib \citep{hunter07} and NumPy \citep{harris20}, whose contributions facilitate the advancement of our work.

\section*{Data Availability}
\label{data_avail}
As extensive simulation datasets for ML begin to emerge \citep{navarro22}, we advocate making these datasets publicly available in order to optimize the benefit from network effects among the research community. Data sharing helps to reduce spending computational resources for data generation, making ML approaches more efficient and sustainable in the long term.

We provide our data, source code, and pre-trained ML models to encourage independent researchers to reproduce our results and to incorporate our methods into their own work. Our data can be accessed at \url{https://phaidra.univie.ac.at/o:1206181}. Our source code and pre-trained ML models can be accessed at \url{https://github.com/littleblacksheep/csv/tree/main}. We kindly ask users to report possible bugs to \href{mailto:winter@ml.jku.at}{winter@ml.jku.at}.

\bibliographystyle{mnras}
\bibliography{citations}

\onecolumn
\appendix
\section{SPH dataset details}
\subsection{Technical details}
\label{SPH_technical}

We use three Nvidia GeForce GTX1080 Ti (11 GB) GPUs for the generation of SPH collision data. The total computation time can be split up into preprocessing, SPH simulations, and postprocessing. Preprocessing and postprocessing are performed on the CPU, whereas SPH simulations are performed on the GPU. The average computation times per simulation (average over 10164 simulations) in these categories are 169\,sec, 42\,min, and 4.8 sec using a single GPU, which resulted in approximately 317 GPU days for the entire SPH dataset. Similar computation times hold for the $N$-body dataset (see \cref{sect:nbody}) from \citet{burger20}.

The relatively large number of datapoints (one datapoint corresponds to one simulation) does not allow for high-resolution runs due to hardware limitations. We perform data reduction on the fly to significantly reduce the memory footprint from approximately 200 TB of raw data. After postprocessing, individual simulation runs require less than 2 MB of storage space. The complete dataset requires about 12.2 GB of storage space.

\begin{figure}
 \includegraphics[width=0.8\columnwidth]{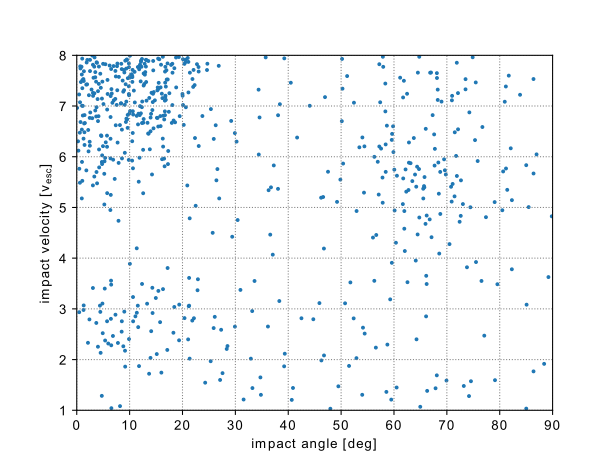}
 \caption{Distribution of invalid simulation runs. The chance that \texttt{miluphcuda} suffers from numerical issues is increased for collision scenarios with high velocities and low impact angles.}
 \label{fig:invalid}
\end{figure}

We performed a total of 10794 SPH simulations. 630 of those runs failed, mostly because they ran into numerical issues either in the setup script or during the actual simulation with \texttt{miluphcuda}. The density of failed runs is higher in the disruption regime compared to other regimes due to excessively small timesteps (via the adaptive time integration). This tends to be more likely in the disruption regime, where very high pressures are more common. We end up with 10164 valid simulations for our dataset. Note that due to the inhomogeneous distribution of invalid simulations, also the distribution of valid simulations is somewhat inhomogeneous, leading to a slightly worse coverage of high-velocity, low-impact-angle scenarios (see \cref{fig:invalid}). We keep configuration files of invalid simulations for possible future data analysis.

\subsection{Pre-collision spin}
\label{rotations}
Let us consider a single rotating body, consisting of $n$ SPH particles. The rotation axis can either be defined in spherical coordinates (radius $r$, azimuth $\phi$, polar angle $\theta$) or in Cartesian coordinates ($x$, $y$, $z$):

\begin{ceqn}
\begin{align}
    x &= r \, \cos(\phi) \, \sin(\theta) \\
    y &= r \, \sin(\phi) \, \sin(\theta) \\
    z &= r \, \cos(\theta)
\end{align}
\end{ceqn}

In our case, $r$ is either the length of the angular momentum vector \textbf{L}, or the rotation period $P_\mathrm{rot}$.

\begin{ceqn}
\begin{align}
    \vec{L}&=\sum_{i=1}^{n} \vec{r}_i \times (m_i \, \vec{v}_i)
\end{align}
\end{ceqn}

Here, $m_i$, $\vec{r}_i$, and $\vec{v}_i$ refer to the masses, positions, and velocities of individual SPH particles w.r.t. the barycenter of the object. The critical rotation period $P_\mathrm{rot, crit}$ is defined such that a test mass at the surface of the (idealized spherical) body is weightless according to Kepler's 3rd law:

\begin{ceqn}
\begin{align}
    P_\mathrm{rot, crit}&=\sqrt{\frac{4\pi^2r^3}{G\, m}}=\frac{2\pi}{\omega_\mathrm{crit}}
\end{align}
\end{ceqn}

Here, $r$, $m$, and $G$ refer to the object radius, its mass, and the gravitational constant. For dataset generation, the rotation speed (angular velocity) $\omega$ is randomly sampled between $\omega=0$ and $\omega=0.2 \times \omega_\mathrm{crit}$, whereas the rotation axis is randomly sampled in Cartesian coordinates.

\section{Machine Learning}
\subsection{Rotation equivariance}
\label{sec:rotation_equivariance}
An optional rotation module can be used as a pre-and postprocessing step. The module rotates the system in the three-dimensional domain, making the ML models equivariant to rotations. Thus, the models are geometrically consistent and do not require to spend their capacity on learning to be rotation equivariant. We propose to rotate and de-rotate the entire system before and after applying learnable modules.

\begin{ceqn}
\begin{align}
    y_i^0 &= rot(\hat{y}_i^0, R_i) \\
    y_i^T &= f(y_i^0, T, \Phi) \\
    \hat{y}_i^T &= rot(y_i^T, R_i^{-1})
\end{align}
\end{ceqn}

$rot$ refers to the rotation module, $f$ is an ML model with its respective learnable parameters $\Phi$, $\hat{y}$ refers to systems with any orientation, whereas $y$ have a fixed orientation. We calculate a rotation matrix $R_i$ for each datapoint such that the system is rotated into a fixed basis. The pre-rotation basis is given via the impact geometry and is calculated as follows (see \cref{fig:collision_geometry}):
\begin{ceqn}
\begin{align}
    \vec{e}_0 = \frac{\vec{v}}{|\vec{v}|} &&
    \vec{e}_1 = \frac{\vec{e}_0 \times \vec{r}}{|\vec{e}_0 \times \vec{r}|} &&
    \vec{e}_2 = \frac{\vec{e}_0 \times \vec{e}_1}{|\vec{e}_0 \times \vec{e}_1|}
\end{align}
\end{ceqn}

$v$ and $r$ are the relative velocity and relative distance vectors between the projectile and the target. The rotation matrix $R_i$ is chosen as the inverse of the pre-rotation basis:
\begin{ceqn}
\begin{equation}
R_i = 
\begin{pmatrix}
e_{00} & e_{01} & e_{02}\\
e_{10} & e_{11} & e_{12}\\
e_{20} & e_{21} & e_{22}
\end{pmatrix}^{-1}
\end{equation}
\end{ceqn}

The rotation module can be easily embedded into ML frameworks for collision treatment.

\begin{figure}
 \includegraphics[scale=0.4]{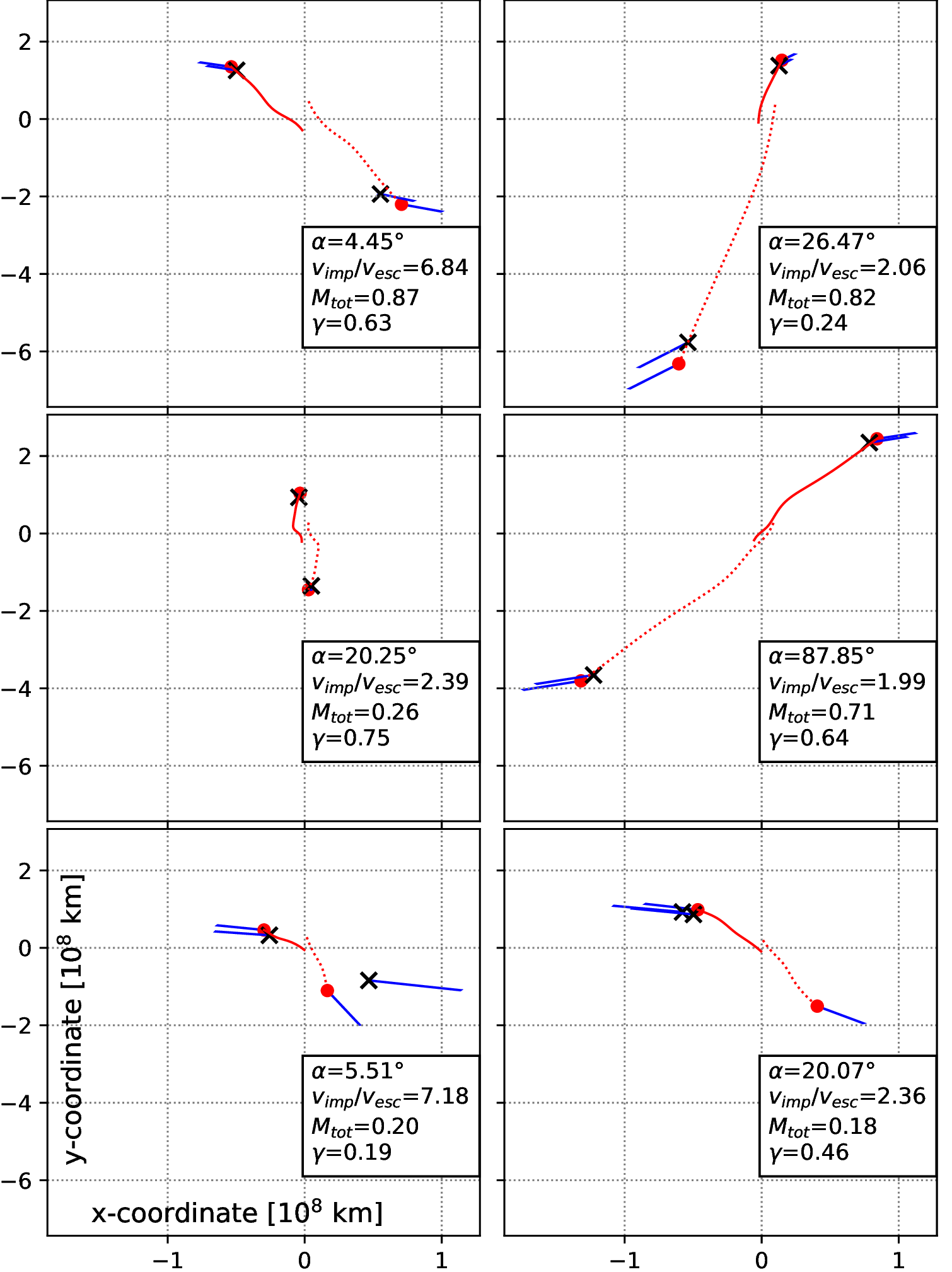} 
 \caption{Visualization of predicted intermediate states (in red) of our RES model for 4 different datapoints of the o.o.d. test set. Each plot depicts two predicted trajectories, where one corresponds to the largest remnant (solid) and the other one to the 2nd-largest remnant (dotted). Final positions are indicated by red dots (predictions) and black crosses (ground truth). Ground truth initial positions are indicated by the starting point of the respective predicted trajectories. Final velocity vectors (for both prediction and ground truth) are indicated by blue arrows for all objects. Note that projectiles are flying into the negative y-direction initially and the x-y plane is the main collision plane. The labels indicate the initial conditions (mass in units of $10^{25}$ kg, cf. \cref{tab:param_space}). We find that z-components (which are caused by rotating bodies) were quite significant in many simulations. Prediction performance in the z-component is comparable to the x -and y components. We observe spatio-temporally continuous transitions from initial to final states, possibly indicating a correlation to the temporal evolution of the physical system to a certain extent. The first 4 panels were cherry-picked for predicted positions in the x-y plane. The last 2 panels visualize typical failure cases of model predictions, where positions and velocities of the 2nd-largest remnants are poorly approximated.}
 \label{fig:trajectories}
\end{figure}

\subsection{Miscellaneous}
\label{sec:hyperparams}

\begin{table}
\begin{center}
\begin{tabular}{ l|l|l } 
 method & description & value \\ 
 \hline
 FFN & number of layers & 8 \\
 FFN & hidden state size & 56\\
 RES & number of steps per simulated hour & 4\\
 RES & number of layers for each module E, R, and D & 3\\
 RES & hidden state size & 64
\end{tabular}
\end{center}
\caption{Model-class hyperparameters for deep learning collision treatment approaches. Hyperparameters are optimized w.r.t. validation performance.}
\label{tab:hyperparams}
\end{table}

\Cref{tab:hyperparams} summarizes optimized hyperparameters for our deep learning models. Formally, model architectures can also be interpreted as hyperparameters.
We tested a handful of different possible architectures and performed ablation studies before settling on the best-performing architecture, which is presented in the main text. For FFN we tested a residual neural network approach and different activation functions. 

We studied LSTMs \citep{hochreiter97, hochreiter98, gers99} in depth and investigated the following setups:
\begin{itemize}
    \item Direct prediction of final states without Euler updates for studying performance of direct predictions versus relative predictions.
    \item LSTM without forget gate for mitigating the vanishing gradient problem.
\end{itemize}

We also investigated several different graph neural network architectures \citep[GNNs;][]{scarselli09, defferrard16, kipf17}. Our best architecture was a latent GNN \citep{dong19} which slightly outperformed the FFN baseline. For the GNN we investigated several architecture variations that use the initial SPH representation in combination with clustering approaches. These variations were mostly inspired by the architecture of \citet{gonzalez20}. Increasing the number of graph nodes may require hierarchical graph structures \citep{martinkus20, ladislav21} to account for learning long-range interactions such as gravity.

For the weight-tied residual network RES we tested state updates via the Bulirsch-Stoer method \citep{stoer80, num_rec92} instead of the classical Euler update, but we did not observe any improvements.

Preliminary investigation of predicted water mass fractions did not allow for in-depth analysis of generalization to the low-water-content regime (i.e., $\zeta_\mathrm{water} < 0.1$), because model predictions were too inaccurate.

We summarize the training and validation performance of our experiments in table \cref{tab:train_and_vali_performance}. Results are consistent with the corresponding test results, which can be found in \cref{tab:exp_main}. 

\begin{table}
\begin{center}
\begin{tabular}{ r|r|r|r|r|r|r|r } 
 method & split & mass & material & position & velocity & total & accuracy \\[0.5ex] 
 \hline
 PIM & training & $0.1388^{+0.0008}_{-0.0012}$& $0.0619^{+0.0002}_{-0.0001}$& $0.4055^{+0.0010}_{-0.0008}$& $0.4411^{+0.0022}_{-0.0018}$& $0.3619^{+0.0010}_{-0.0007}$& $0.1667^{+0.0000}_{+0.0000}$\\[0.5ex]
 LIN & training & $0.0441^{+0.0003}_{-0.0003}$& $0.0280^{+0.0001}_{-0.0001}$& $0.1617^{+0.0005}_{-0.0006}$& $0.1579^{+0.0011}_{-0.0015}$& $0.1375^{+0.0006}_{-0.0007}$& $0.2296^{+0.0068}_{-0.0088}$\\[0.5ex]
 FFN & training & $0.0060^{+0.0003}_{-0.0004}$& $0.0115^{+0.0002}_{-0.0001}$& $0.0403^{+0.0007}_{-0.0007}$& $0.0446^{+0.0006}_{-0.0007}$& $0.0367^{+0.0005}_{-0.0005}$& $0.4422^{+0.0444}_{-0.0416}$\\[0.5ex]
 RES & training & \bm{$^*0.0047^{+0.0001}_{-0.0001}$}& \bm{$^*0.0108^{+0.0002}_{-0.0003}$}& \bm{$^*0.0272^{+0.0002}_{-0.0003}$}& \bm{$^*0.0339^{+0.0003}_{-0.0006}$}& \bm{$^*0.0272^{+0.0002}_{-0.0002}$}& \bm{$^*0.5234^{+0.0089}_{-0.0182}$}\\[0.5ex]
 \hline
 PIM & validation & $0.1388^{+0.0048}_{-0.0032}$& $0.0619^{+0.0004}_{-0.0008}$& $0.4056^{+0.0031}_{-0.0041}$& $0.4411^{+0.0071}_{-0.0087}$& $0.3619^{+0.0030}_{-0.0041}$& $0.1667^{+0.0000}_{+0.0000}$\\[0.5ex]
 LIN & validation & $0.0443^{+0.0012}_{-0.0009}$& $0.0280^{+0.0005}_{-0.0008}$& $0.1619^{+0.0024}_{-0.0018}$& $0.1581^{+0.0066}_{-0.0042}$& $0.1377^{+0.0033}_{-0.0021}$& $0.2380^{+0.0485}_{-0.0220}$\\[0.5ex]
 FFN & validation & $0.0066^{+0.0003}_{-0.0004}$& $0.0117^{+0.0002}_{-0.0001}$& $0.0420^{+0.0016}_{-0.0017}$& $0.0458^{+0.0023}_{-0.0032}$& $0.0379^{+0.0012}_{-0.0021}$& $0.4558^{+0.0328}_{-0.0334}$\\[0.5ex]
 RES & validation & \bm{$^*0.0052^{+0.0003}_{-0.0004}$}& \bm{$^*0.0111^{+0.0003}_{-0.0004}$}& \bm{$^*0.0320^{+0.0019}_{-0.0012}$}& \bm{$^*0.0382^{+0.0015}_{-0.0022}$}& \bm{$^*0.0309^{+0.0014}_{-0.0010}$}& \bm{$^*0.5381^{+0.0438}_{-0.0631}$}\\[0.5ex]
 \hline
 PIM & training, Timpe & $0.2080^{+0.0003}_{-0.0003}$& $0.1933^{+0.0014}_{-0.0009}$& $0.3598^{+0.0019}_{-0.0014}$& $0.2897^{+0.0017}_{-0.0017}$& - &$0.1667^{+0.0000}_{+0.0000}$\\[0.5ex]
 LIN & training, Timpe & $0.0528^{+0.0005}_{-0.0004}$& $0.0493^{+0.0004}_{-0.0003}$& $0.2040^{+0.0013}_{-0.0014}$& $0.1691^{+0.0009}_{-0.0007}$& - &$0.3019^{+0.0063}_{-0.0037}$ \\[0.5ex]
 FFN & training, Timpe & $0.0124^{+0.0003}_{-0.0002}$& $0.0145^{+0.0002}_{-0.0002}$& $0.0757^{+0.0017}_{-0.0009}$& $0.0589^{+0.0002}_{-0.0003}$& - &$0.4172^{+0.0040}_{-0.0040}$ \\[0.5ex]
 RES & training, Timpe & \bm{$^*0.0112^{+0.0004}_{-0.0004}$}& \bm{$^*0.0138^{+0.0005}_{-0.0006}$}& \bm{$^*0.0658^{+0.0010}_{-0.0008}$}& \bm{$^*0.0548^{+0.0007}_{-0.0007}$}& - &\bm{$^*0.4252^{+0.0063}_{-0.0059}$}\\[0.5ex]
 \hline
 PIM & validation, Timpe & $0.2080^{+0.0013}_{-0.0011}$& $0.1933^{+0.0035}_{-0.0058}$& $0.3598^{+0.0056}_{-0.0077}$& $0.2897^{+0.0066}_{-0.0070}$& - &$0.1667^{+0.0000}_{+0.0000}$\\[0.5ex]
 LIN & validation, Timpe & $0.0529^{+0.0015}_{-0.0009}$& $0.0495^{+0.0014}_{-0.0005}$& $0.2044^{+0.0046}_{-0.0064}$& $0.1695^{+0.0043}_{-0.0046}$& - &$0.3010^{+0.0181}_{-0.0105}$\\[0.5ex]
 FFN & validation, Timpe & $0.0137^{+0.0010}_{-0.0008}$& $0.0155^{+0.0009}_{-0.0006}$& $0.0834^{+0.0056}_{-0.0025}$& \bm{$0.0642^{+0.0024}_{-0.0019}$}& - &$0.4196^{+0.0163}_{-0.0116}$\\[0.5ex]
 RES & validation, Timpe & \bm{$^*0.0130^{+0.0013}_{-0.0011}$}& \bm{$0.0152^{+0.0005}_{-0.0003}$}& \bm{$0.0826^{+0.0054}_{-0.0035}$}& $0.0649^{+0.0028}_{-0.0017}$& - &\bm{$0.4265^{+0.0059}_{-0.0049}$}
\end{tabular}
\end{center}
\caption{Training and validation performance (RMSE and balanced accuracy, both measured on the final state) of different approaches for planetary collision treatment. Classification is performed as a postprocessing step on top of predicted masses. We assume the SPH simulation data to be the ground truth. For single-task learning (lower half of the table), each entry corresponds to the performance of individually trained ML models and column headers indicate optimized tasks, respectively. We use data from \citet{timpe20} to obtain results in the lower half. Best results are indicated in bold, whereas * indicate statistically significant results according to a Wilcoxon test (comparing FFN and RES). Our proposed RES model outperforms the other baseline methods in most experiments.}
\label{tab:train_and_vali_performance}
\end{table}

We provide pre-trained FFN and RES models for direct integration into existing $N$-body frameworks. We train these models using the entire development set (i.e., training + validation data) and report their performance in \cref{tab:pretrain_performance}.

\begin{table}
\begin{center}
\begin{tabular}{ r|r|r|r|r|r|r|r } 
 method & split & mass & material & position & velocity & total & accuracy \\ 
 \hline
 FFN & development & $0.0055$& $0.0111$& $0.0399$& $0.0442$& $0.0364$& $0.4535$\\
 RES & development & $0.0046$& $0.0108$& $0.0273$& $0.0341$& $0.0273$& $0.5035$\\
 \hline
 FFN & o.o.d. test & $0.0119$& $0.0120$& $0.0483$& $0.0432$& $0.0405$& $0.5593$\\
 RES & o.o.d. test & $0.0104$& $0.0125$& $0.0375$& $0.0425$& $0.0357$& $0.5091$
\end{tabular}
\end{center}
\caption{Performance of provided pre-trained models.}
\label{tab:pretrain_performance}
\end{table}

\subsection{Incorporation of ML models into $N$-body frameworks}
\label{sec:drop_in}
All our models can in principle be employed for collision treatment within existing $N$-body simulation frameworks. In the following we list some important aspects related to implementation and technical details thereof:

\begin{itemize}
    \item As our data is highly imbalanced (see \cref{fig:stewart} and \cref{tab:classes}), different use cases might require re-training the ML model using different data subsets. This benefit might be especially true for classification accuracy, which is sensitive to class-imbalances. Note that using the entire dataset might lead to miss-classifications of collision outcome types (e.g. compare \cref{fig:stewart} vs. \cref{fig:post_classes}). Also, depending on the use case, one might want to incorporate additional restrictions (such as imposing conservation laws) as post-processing step on top of model predictions. Further details are elaborated in \cref{sec:regression}.
    \item Our models require consistent input features, produced by collisions that lie within our parameter space as defined in \cref{tab:param_space}. Having consistent input features also includes applying the identical preprocessing pipeline (e.g. the numerical scaling of features) in analogy to the preprocessing pipeline that is used for model training. We believe that producing consistent input features is the most error-prone process when implementing ML methods for collision treatment. For sophisticated frameworks or if o.o.d. datapoints are expected to be common, one may even consider applying anomaly detection methods to check for potential invalid inputs.
    \item Note that we define the onset of the collision process before objects come in direct contact in order to include tidal effects. This should be considered for triggering collision events in $N$-body frameworks. We propose taking two measurements, one when tidal interaction begins, and another one once the objects come into direct contact (cf. \cref{fig:collision_geometry}). We initialize the colliding bodies at a distance of $d_\mathrm{initial} = f_i \times (R_t + R_p)$. $R_t$ and $R_p$ are the target and projectile radii, while the initial distance factor $f_i$ is a hyperparameter (between $f_i=3$ and $f_i=7$ for our data). Therefore, reasonable estimates of $R_t$ and $R_p$ might be required to define the first measurement in $N$-body simulations.
    \item Collisions can happen at arbitrary orientations. We provide a rotation module to account for rotation-equivariance of our ML models. We recommend re-training ML models using data augmentation (i.e., randomly rotating the systems) in combination with the rotation module as a sanity check for guaranteeing rotation-equivariance. Note that all ML models in this paper were trained without using the rotation module, i.e., projectiles flying into the negative y-direction initially, where the main collision plane is the x-y plane. Our data preprocessing is also specifically adjusted for this fixed-orientation setup, i.e., different axes getting scaled differently in order to account for the variability in the respective directions.
    \item Our ML models can be ran on CPUs if using GPUs is unfeasible or none are available. We estimate that CPU inference times $\tau_\mathrm{i, ML}$ are on the order of 1-10\,s.
    \item As discussed in Sect.~\ref{sect:ml_exp_setup}, extracting reliable post-collision rotation states from our data remains an open issue, potentially restricting the full usability of our ML models in N-body simulations, specifically for full tracking of rotation states over several collisions. However, once post-collision rotation states can be extracted reliably from collision simulations, they can be easily incorporated into our multi-task regression objective.
\end{itemize}

\bsp 
\label{lastpage}
\end{document}